\documentclass{aa}
\usepackage{graphicx}
\usepackage{float}
\usepackage{gensymb}
\usepackage{txfonts}
\usepackage{subfigure}
\usepackage{float}
\usepackage{comment}
\usepackage{multirow}
\usepackage{booktabs}
\usepackage{soul}
\usepackage{comment}
\usepackage[breaklinks, colorlinks, citecolor=blue, linkcolor=blue]{hyperref}
\newcommand{\RN}[1]{%
  \textup{\uppercase\expandafter{\romannumeral#1}}%
}
\usepackage{xcolor}

\newcommand\vertwo[1]{#1}

\begin{document} 

   \title{Simulating Rayleigh-Taylor induced magnetohydrodynamic turbulence in prominences}

%   \subtitle{I. Overviewing the $\kappa$-mechanism}
\author{M. Changmai \inst{1}
	    \thanks{\email{madhurjya.changmai@kuleuven.be}}
	    \and J. M. Jenkins\inst{1}
	    \and J. B. Durrive\inst{1}
	    \and R. Keppens\inst{1}
	    }

\institute{Centre for mathematical Plasma-Astrophysics, Celestijnenlaan 200B, 3001 Leuven, KU Leuven, Belgium}

   \date{}

% \abstract{}{}{}{}{} 
% 5 {} token are mandatory
 
  \abstract
   {}
  {Solar prominences represent large-scale condensations suspended against gravity within the solar atmosphere. The Rayleigh-Taylor (RT) instability is proposed to be one of the important fundamental processes leading to the generation of dynamics at many spatial and temporal scales within these long-lived, cool, and dense structures amongst the solar corona. We aim to study such turbulent processes using high-resolution, direct numerical simulations of solar prominences.}
  {We run 2.5D ideal magnetohydrodynamic (MHD) simulations with the open-source {\tt MPI-AMRVAC} code far into the nonlinear evolution of an RT instability perturbed at the prominence-corona interface. Our simulation achieves a resolution down to $\sim 23$~km on a 2D $(x,y)$ domain of size 30~Mm~$\times$~30~Mm. We follow the instability transitioning from a multi-mode linear perturbation to its nonlinear, fully turbulent state. Over the succeeding $\sim 25$~minute period, we perform a statistical analysis of the prominence at a cadence of $\sim 0.858$~s.}
  {We find the dominant guiding $B_z$ component induces coherent structure formation predominantly in the vertical velocity $V_y$ component, consistent with observations, demonstrating an anisotropic turbulence state within our prominence. We find power-law scalings in the inertial range for the velocity, magnetic, and temperature fields. The presence of intermittency is evident from the probability density functions of the field fluctuations, which depart from Gaussianity as we consider smaller and smaller scales. In exact agreement, the higher-order structure functions quantify the multifractality, in addition to different scale characteristics and behavior between the longitudinal and transverse directions. Thus, the statistics remain consistent with the conclusions from previous observational studies, enabling us to directly relate the RT instability to the turbulent characteristics found within quiescent prominence.}
   {}


   \maketitle
%
%-------------------------------------------------------------------

\section{Introduction}

Solar prominences are large structures containing relatively cool ($10^{4}$~K) and dense condensations embedded within the far hotter ($10^6$~K) and the tenuous solar corona. The characteristic dimensions of prominences range from $60$ to $600$~Mm in length, $15$ to $100$~Mm in height, and $5$ to $15$~Mm in thickness \citep{1995ASSL..199.....T}. They are mainly classified into three types: active, intermediate, and quiescent, based on their lifetime, location on the Sun, and strength of the interrelated magnetic field \citep{mackay2010physics}. They originate from filament channels and develop above magnetic polarity inversion lines (the observed delineation between predominantly positive and negative fields at photospheric heights). Quiescent prominences are usually found in quiet regions in the Sun's atmosphere, at a range of latitudes, and populate the largest of the aforementioned scales. 

A typical quiescent prominence exists within quiet regions of the solar corona with an electron number density $\sim 10^{11}$~cm$^{-3}$ \citep{1986NASCP2442..149H} and temperatures around $\sim 10^4$~K \citep{1995ASSL..199.....T}. According to early observations, the internal magnetic field is mostly horizontal to the solar surface, making an acute angle of about 40\degree\ with respect to the long axis of the prominence \citep{1998ASPC..150..434B}. \vertwo{With the exception of \citet{2006ApJ...642..554M}, whom found a nearly vertical field orientation, the majority of subsequent works continue to find evidence of a horizontally-oriented internal magnetic field of magnitude a few tens of Gauss and a slightly broader range of thread deviation angles ($>15$\degree) from the guiding axis \citep[][]{2003ASPC..307..468C, 2003ApJ...598L..67C, 2016ApJ...826..164L}.} 

Initially, \citet{1967ApJ...150..313R} estimated the plasma~$\beta$ ratio, thermal to magnetic pressure, to be $\approx 0.02$. Such a low $\beta$ in prominences means that the Lorentz force, particularly the magnetic tension component, must be responsible for supporting the prominence material; the key force balance for such a magnetohydrostatic equilibrium then is magnetic tension and gravity. \cite{2021PhyD..42332838Z} discussed the relative importance of the key forces in prominence dynamics. Most notably, gravity can be dynamically essential in prominences even for $\beta < 1$, if the system size is larger than the pressure scale height. \vertwo{For quiescent prominences, more recent studies have found the pressure and magnetic field strength values to be of the order $\approx 0.1$~dyne~cm$^{-2}$ and $\,10$~G, respectively, increasing to as much as $0.4$\,--\,$0.6$~dyne~cm$^{-2}$ and $60$\,--\,$80$~G \citep[for both quiescent and active environments,][]{1986NASCP2442..149H, 2003ApJ...598L..67C, 2018ApJS..236...35L, 2019A&A...631A.146S}.} Hence, since the initial plasma beta estimates of \citet{1967ApJ...150..313R}, this further research enables us to conclude that the range plasma $\beta$ within prominences can be across four orders of magnitude, ranging between $0.0004\,$--$\,4$. Naturally, this has large implications for the range of plasma dynamics to be found within solar prominences.

Detailed observations of these phenomena found a spatio-temporal range of internal dynamics, curiously dominated by many vertical striations \citep{Berger_2008}. These prominence plumes are observed as dark, rising voids in cool spectral lines. The upflowing plumes measured by the Hinode Solar Optical Telescope (SOT) observations by \citet{Berger_2010} were found to have characteristic initial widths of 0.5 to 1.7~Mm. These structures appear to rise with near-constant speeds of 10 to 17~km~s$^{-1}$, areal dimensions between 2 to 6~Mm, reach a maximum height of 11 to 17~Mm, before eventually descending as pillars with an acceleration of 0.48 to 3.2~km~s$^{-2}$. \citet{Berger_2010}, \citet{2010SoPh..267...75R}, \citet{2012A&A...540L..10I}, and \citet{2019ApJ...874...57M} propose that the RT instability may be a key mechanism for driving upflow plumes and descending pillars. RT instability is a fundamental instability driven by gravity where the heavier fluid is accelerated against a lighter fluid separated by an interface that is unstable to perturbations. In an ideal, stratified magnetohydrodynamic (MHD) model, a discontinuous interface between the heavier and lighter fluids is similarly unstable to such perturbations. Herein, the deposition of hydrodynamic baroclinic vorticity amplifies initial deformations of the interface as a result of the misalignment between the local pressure and density gradients \citep{2021PhyD..42332838Z}, where the linear growth rate then increases with increasing mode number \citep{chandrasekhar1961international}.

The high-resolution observations from Hinode SOT are largely responsible for our understanding of the non-linear behavior of their internal dynamics \citep[][]{chae2010dynamics,hillier2012numerical}. \citet{chae2010dynamics} found from the observations that the vertical fine substructures, termed knots, were impulsively accelerated in the downward direction. The authors related this to magnetic reconnection and the interchange of the magnetic configuration. The SOT observations of \citet[][]{Berger_2008,Berger_2010,berger2011magneto} additionally linked these dynamics to the aforementioned magnetic RT instability, as well as the well-known Kelvin-Helmholtz (KH) instability, subsequently validated in numerical works \citep[][]{2011ApJ...736L...1H,hillier2012numerical}. Naturally, these earlier studies focused on the velocity field, specifically this wide range of motions in the vertical direction \citep[][]{1981SoPh...70..315E,kubota1986vertical} and highlighted, importantly, how these upflows subsequently evolve into vortices \citep{liggett1984rotation}. Furthermore, the upward flows can be supersonic, and evidence of bow-shock compression is reported by \citet{Berger_2010}. And so, in general, quiescent prominences are seen to exhibit spatio-temporal evolution characterized by high variability. The dynamic nature and the large prevailing Reynolds numbers then indicate the fluctuations present within prominences to be turbulent in nature. 

Following such conclusions, \citet{leonardis2012turbulent} used the Ca~{\sc ii}~H intensity images from Hinode SOT to explore the quantitative signatures of turbulence related to solar prominences for the first time. The authors looked at the turbulence non-Gaussianity, multifractality, and intermittency aspects, studying the power-law scalings within the power spectral density (PSD) as a function of wavenumber. From this, \citet{freed2016analysis} also used Hinode SOT data to investigate the characteristics of the plasma flow, focusing primarily on the kinetic energy and vorticity. Shortly after that, \citet{hillier2017investigating} used $H\alpha$ Dopplergrams from  Hinode SOT to investigate the nature of turbulent prominence motions, applying structure function (SF) analysis to the velocity increments. In this paper, we will use direct numerical simulations and analyze our MHD model of a quiescent prominence in much the same fashion as used in these previous studies.

Theoretical studies using ideal MHD simulations have been carried out by \citet[][]{hillier2012numerical, hillier2012numerical2,keppens2015solar,xia2016formation}, wherein the authors were able to reproduce upflows and downflows with speeds consistent with observational values. \citet[][]{hillier2012numerical, hillier2012numerical2} performed 3D MHD simulations of the RT instability mode development in the lower regions of prominence boundaries to investigate the non-linear stability of the Kippenhahn-Schlüter model \citep{1957ZA.....43...36K}. A total of $40 \times 150 \times 320 \, (x \times y \times z)$ grid points were used in this 3D numerical simulation. The nonlinear RT instability formed low-density filamentary structures aligned in the magnetic-field direction, which was important for creating upflows in the prominence. In \citet{keppens2015solar}, the macroscopic behavior of the RT instability and its ability to form bubbles and pillars were studied using 3D ideal-MHD simulations. In this study, a resolution of $600 \times 600 \times 280$ was achieved using dynamic grid refinement to simulate the late nonlinear stages for a weak field case of 8~G. The study shows the development of substructures along the edges of the largest bubbles in the system in line with the existence of strong shear flows previously noted within observations. These studies mainly focused on the evolution of the early (up to a saturation regime) nonlinear RT stages in a quiescent prominence. Still, the study of the ensuing, persistent turbulence has been limited so far. 

Continuing, \citet{2015ApJ...799...94T} used 3D MHD simulations of global prominence models to investigate and relate the prominence morphology and dynamics to the different model parameters. Investigating the time evolution of solar prominences embedded in sheared magnetic arcades, the authors demonstrated that magnetic shear and low plasma $\beta$ stabilize the configuration. The \citet{2015ApJ...799...94T} study was performed using a resolution of $150 \times 150 \times 100$. The setup had no initial velocity perturbation, but due to ad-hoc mass deposition, the system generated vertical velocity perturbations as mass got pulled down by gravity. The flows in \citet{2015ApJ...799...94T} are due to gravitational potential energy leading to the formation of RT instability-like dynamics; the coarse resolution prevented direct comparisons against observed scales. Using 3D MHD simulations, \citet{2016ApJ...820..125T} then continued with a study of the temporal evolution of a cold solar prominence embedded in a three-dimensional magnetic flux rope where the orientation of the prominence along the flux rope axis prevented the development of the RT instability altogether. 

Since then, \citet{2016ApJ...825L..29X} used 3D MHD simulations to investigate the non-linear magnetoconvective motions within a twin-layer prominence beginning with a predominantly horizontal magnetic field and a gravitationally force-balanced solar atmosphere. To do so, a box of $30 \times 30 \times 30$~Mm having a grid resolution of $600 \times 600 \times 600$ with the smallest grid cell size of 50~km was taken. The initial evolution of the two layers is very similar due to the supporting magnetic field, which becomes interchanged due to the RT instability. This similarity decreases in the later turbulent phase, and as a collective kink deformation was induced in the twin-layer prominence. \citet{2018ApJ...869..136K} performed high-resolution 3D MHD modeling of the RT instability in prominences, which was able to reproduce magnetic RT dynamics within a prominence formed \textit{pseudo-ab-initio} via the reconnection–condensation model. As an initial condition, the necessary perturbations were added to density and velocity in this study, resulting in a heterogenous radiative condensation. Interestingly, the condensation rate found was
comparable to the mass drainage rate to maintain the total mass of the prominence. Compared to \citet{keppens2015solar}, where the domain had a transition region above the lower chromosphere boundary, which resulted in the reflection of material from falling spikes due to RT instability to then rise to coronal heights, the \citet{2018ApJ...869..136K} used a full coronal domain. For a more detailed explanation of RT instability in prominences, the reader is directed to \citet{2018RvMPP...2....1H}, and similarly to \citet{Jenkins:2022} wherein the authors advocate for a timely migration towards generality through the gravitational interchange instability.

In summary, each mentioned study focused mostly on the early linear and nonlinear stages of RT instability in prominences using a wide array of models differing significantly in complexity. To clearly understand whether the RT instability could be a key factor leading to the turbulence observed in prominences, a physically detailed and long-duration run of the simulation is needed. The derived statistics should then be compared directly to the observations. Furthermore, and based on the results presented by all of these authors, it is clear that a yet-higher resolution is required to capture the smaller scales of the instability and any resulting turbulence, without which the dynamics at these scales will be dominated by numerical diffusion. The RT instability growth will be limited to the earlier, larger-scale evolutions. 

Analytical and numerical studies of the RT instability in solar prominences are mostly carried out using the single-fluid MHD approach. A limited number of numerical studies using the ambipolar term to characterize partial ionization effects have been completed for flux emergence (\citealp{2007ApJ...666..541A,2015RSPTA.37340268M}) and for prominences (\citealp{2014A&A...564A..97D,2014A&A...565A..45K}). In the context of even more complete two-fluid approaches, numerical work regarding solar prominences are far more scarce. \citet{2012ApJ...754...41D} used a two-fluid approach to study the linear onset and growth rate of the RT instability within the partially ionized plasma. Recent works of \citet{2021A&A...646A..93P,2021arXiv210112731P} have used a non-linear two-fluid model to study the effects of partial ionization on the growth rate of RT instability. The authors find the deviation from MHD to be important only at extremely small scales around $\sim 3$ km. The ambipolar term and true two-fluid treatments indeed affect the growth of small scales. Still, we here will ignore these effects of plasma-neutral couplings in our simulation, which can be a topic for future investigation. In our study using the ideal MHD approach, we will deal with scales above 20 km to understand the scale dependence on the turbulence previously observed in solar prominences due to the RT instability. At those scales, a single fluid MHD description suffices.

This work presents the first study of high-resolution simulated turbulence initiated by the RT instability within a 2.5D solar prominence \vertwo{\citep[cf.][]{2021arXiv211213043P}}. We have divided our paper into two sections. First, in Section~\ref{sec-setup}, we discuss how we set up the numerical simulation, that is, the governing equations, initial, and boundary conditions to represent a 2.5D ideal MHD prominence unstable to the RT instability. We discuss the evolution of this instability from a linear phase to a non-linear phase and finally to a turbulent phase where the energy transfer through the intermediate scales takes place. Then, in Section~\ref{sec-ana}, we analyze the turbulence in our simulated prominences with different field strengths and discuss the statistical methods that help us understand the prominence turbulence features. Finally, we present our summary and conclusions in Section~\ref{sec-conc}.

\section{Numerical setup}\label{sec-setup}

\begin{figure}
	\centering
	\resizebox{\hsize}{!}{\includegraphics{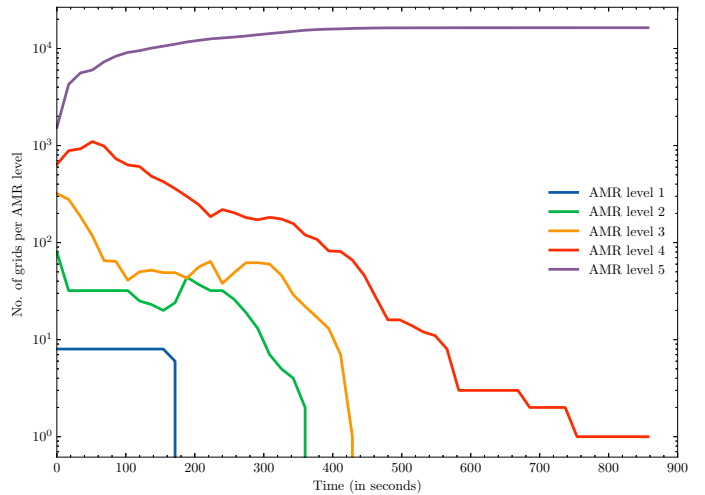}}
	\caption{Number of grids per AMR level during the evolution of the flow. }
	\label{amr_level}
\end{figure}

To investigate the characteristics of turbulence induced due to the RT instability, we perform 2.5D high-resolution ideal MHD simulations of a quiescent prominence in the solar atmosphere using the parallelized, open-source Adaptive Mesh Refinement Versatile Advection Code or {\tt MPI-AMRVAC} \citep{keppens2012parallel,porth2014mpi,xia2018mpi,keppens2020mpi}. It is a parallel adaptive mesh refinement code that solves (primarily hyperbolic) partial differential equations using conservative variables (density, momentum density, total energy density, and magnetic field) across a finite volume grid. We choose a 2.5D setup to simplify the problem analytically and computationally, representing an efficient way of resolving the small scales for the turbulent study of solar prominences using the ideal MHD model. The 2.5D nature implies that we have velocity and magnetic field components perpendicular to the simulated $(x,y)$ plane, and invariance is assumed along this $z$-direction. We set up a large simulation domain using cartesian geometry, spanning $30$~Mm horizontally along the solar surface and $30$~Mm in height extending from the photosphere\,--\,low chromosphere into the solar corona. We adopt a base grid resolution of $40 \times 40$ and use a maximum refinement level of six to achieve a final grid resolution of $1280 \times 1280$. This is represented by the time evolution of the AMR grid level in Figure~\ref{amr_level}, where the number of grids per AMR level is shown. It starts from the base grid AMR level 0 and goes until the highest AMR level 5. \vertwo{This Figure shows an order of magnitude change in the coverage at the highest (most expensive) AMR level, which translates into an instant corresponding reduction in compute time. The final 600 seconds or so then cost exactly the same as a uniform grid resolution run.} As we intend to study and quantify turbulence in a dynamically evolving, stratified prominence setup, the use of adaptive mesh refinement (AMR) allows efficient computation of the transition to a more volume-filling turbulent state and saves \vertwo{appreciable} computing resources compared to employing a uniform mesh at $1280 \times 1280$ \vertwo{at all times}. This is characterized by the highest level grid domain coverage percentage, which climbs from 59.44 \% at $t=0$~s to nearly 100 \% by $t=420$~s.

We note that the observational study of \citet{leonardis2012turbulent} on the turbulent characteristics of intensity fluctuations in prominences used intensity images obtained from the Hinode spacecraft, which gave a spatial resolution of $\sim 77.22$~km with an average temporal resolution of $16.8$~s for a total time interval of $\sim 4.5$~hours. Each image consisted of $800 \times 400$~pixels. We achieve a higher spatial resolution of $\sim 23$~km and base our analysis on data with a temporal resolution of $0.858$~s for a total time interval of $\sim 25$~minutes.

% \vspace{0.5cm}
\subsection{Governing equations}

We solve the ideal MHD equations of conservation of mass, momentum, total energy density, and induction by using the {\tt MPI-AMRVAC} code. The equations are given by,
\begin{eqnarray}
\label{eqn:idealmhd1}
	\partial_t \rho + \nabla \cdot (\textbf{v} \rho) = 0 \,,\\
\label{eqn:idealmhd2}	\partial_t(\rho \textbf{v}) + \nabla \cdot (\textbf{v}\rho \textbf{v} - \textbf{BB}) + \nabla p_{tot} = \rho \textbf{g} \,,\\
\label{eqn:idealmhd3}	\partial_t e + \nabla \cdot (\textbf{v}e - \textbf{BB}\cdot \textbf{v} + \textbf{v}p_{tot}) = \rho \textbf{g}\cdot\textbf{v}\,,\\
\label{eqn:idealmhd4}	\partial_t \textbf{B} + \nabla \cdot (\textbf{vB} - \textbf{Bv}) = 0\,,
\end{eqnarray}
where $p = (\gamma - 1)(e - \rho \textbf{v}^2/2 - \textbf{B}^2/2)$ and $p_{tot} = p + \textbf{B}^2/2$, and the quantities $\rho$, $\textbf{v}$, $p$, $p_{tot}$, $e$, $\gamma$ and $\textbf{B}$ denote density, velocity vector, plasma pressure, total pressure, total energy density, adiabatic index $\gamma$ is the ratio of specific heats, taken as 5/3 under the assumption of a monoatomic ideal gas, and magnetic field vector respectively. The latter is measured in units for which magnetic permeability equals $\mu_0 = 1$. The plasma temperature $T$ is defined with the ideal gas law,
\begin{equation}
    p \mu = R \rho T \,,
\end{equation}
where $R$ and $\mu$ are the gas constant and mean molecular mass, respectively. We set up the following normalization quantities of unit length $= 1 \times 10^9$~cm, unit temperature $= 1 \times 10^6$~K, unit number density $= 1 \times 10^9$~cm$^{-3}$ in our code. Equivalently in code units, the unit velocity is $1.16 \times 10^7$~cm~s$^{-1}$, the unit mass is $2.34 \times 10^{12}$~g, the unit density is $2.34 \times 10^{-15}$~g~cm$^{-3}$, and the unit time is $85.87$~s. The constant gravitational acceleration is then formulated as dimensionless according to the magnitude at the solar surface, $= - 2.74 \times 10^4 \frac{\mbox{unit length}}{\mbox{unit velocity}^2}$, in the $y$-direction.

The Eqs.~(\ref{eqn:idealmhd1})-(\ref{eqn:idealmhd4}) are solved on an evolving, hierarchical block-adaptive grid, using a four$-$stage, third$-$order Runge-Kutta method \citep{ruuth2002two}, with the HLLD Harten-Lax-van-Leer approximate Riemann Solver for multiple discontinuities \citep{miyoshi2005multi} and using a third-order asymmetric slope limiter given by \citet{koren1993robust}. The HLLD method is a positivity-preserving method designed to resolve high-resolution MHD evolutions more robustly and gives higher efficiency than linearized Riemann solvers with respect to numerical accuracy. To ensure we achieve a stable state of turbulence, we use refinement criteria based on instantaneous plasma properties and their gradients, at the current time step following Lohner's prescription. We use the GLM method \citep{2002JCoPh.175..645D} for the magnetic field divergence fix, where the divergence constraint is coupled with the conservation laws by introducing an additional scalar variable $\psi$. We adopt a Courant value of 0.8 to ensure a stable time step within the explicit time integration of each iteration. 

\vspace{0.5cm}
\subsection{Initial and boundary conditions}

The $2$D simulation domain is initialized at $t=0$ on $0 \le x \le 30$~Mm and $0 \le y \le 30$~Mm. The bottom of the initial, purely inserted prominence lies at $y_b = 11.25$~Mm, and the prominence transitions to the outer corona starting at $y_p = 17.5$~Mm. The mean magnetic field strength is taken to be around 4~G. The initial magnetic field is taken as $\textbf{B} = (0.1G, 0, B_z(y))$ which is a purely 1D stratified profile, taken to be non-uniform with an exponential decrease of its strongest component $B_z(y)$ between a height of $y_b$ and $y_p$. The magnetic field makes an angle ($\theta$) of $\sim$ $88.5\degree$ with the horizontal $x$-axis, so it is nearly perpendicular to the simulated plane. The exact analytic form of $B_z(y)$ is given by,
\begin{equation}
B_z = \begin{cases}
B_{z0}, & y < y_b\\
B_{z0} \exp \Bigg[- \left(y-y_b\right)/ \lambda_B\Bigg], & y_b \leqslant y \leqslant y_p\\
B_{z0} \exp \Bigg[- \left(y_p-y_b\right)/ \lambda_B\Bigg], & y_p < y ,
\end{cases}
\end{equation}
where the parameters are $B_{z0} = 4$~G and pressure scale-height, $\lambda_B = 15$~Mm. Within the transition region between $y_b$ and $y_p$, the Lorentz force $((\nabla \times \textbf{B}) \times \textbf{B})$ is not zero which leads to the magnetic field configuration in the region to be non force-free. In the presence of gravity, the governing equation for magnetostatic equilibrium becomes,
\begin{equation}
- \nabla p + (\nabla \times \textbf{B}) \times \textbf{B} + \rho \textbf{g} = 0, \label{eq:mhsBalance}
\end{equation}
which constitutes the vertical force balance between the pressure, Lorentz force, and gravity of the plasma in the prominence material. A decrease in the magnetic field strength within the body of the prominence establishes an upward magnetic pressure force holding the prominence material against the gravity force in the solar atmosphere while also inducing a continuous shear due to the rotating magnetic field. 

\begin{figure}
	\centering
 	\resizebox{\hsize}{!}{\includegraphics{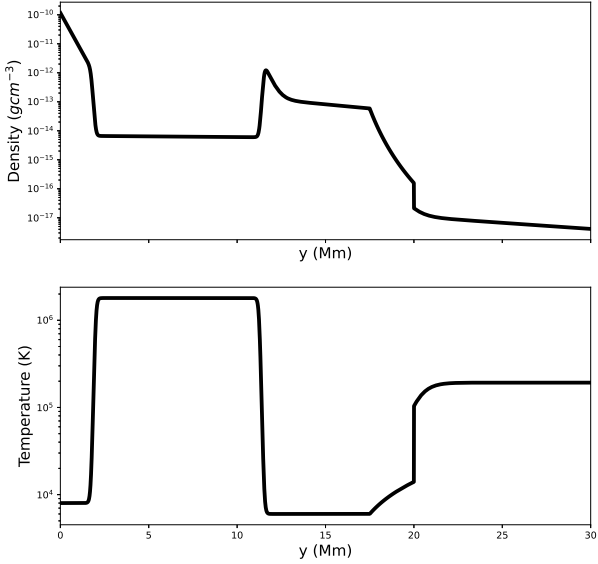}}
	\caption{The derived density (top) and enforced initial temperature (bottom) profiles in the vertical direction at the initial state. The temperature is set to be 8000 K in the chromosphere, and transitions to 1.8 $\times 10^6$ K for the corona, dropping at $y=y_b$ to 6000 K for the lower part of the prominence and then a linear increasing profile from 6000-14000 K at $y = y_p$ to $y = 20$ Mm for the upper part of the prominence.}
	\label{init_temp}
\end{figure}

The initial vertical equilibrium, including the chromosphere, a coronal part below the prominence, the prominence, and the corona above, is thus set up by first initializing a magnetohydrostatic equilibrium from the vertical temperature profile as shown in Figure~\ref{init_temp} and given by,
\begin{align}
&T_j =\nonumber\\ &\begin{cases}
T_{ch} + \frac{1}{2}\left( T_{co} - T_{ch}\right)\left( \tanh{\frac{y_j - 2}{0.1}+1}\right), & y_j \leqslant 4 \\
T_{co} + \frac{1}{2}\left( T_{co} - T_{promin}\right)\left( \tanh{\frac{y_j - 11.25}{0.1}+1}\right), & 4 < y_j < 17.5\\
T_{promin} + (T_{promax}-T_{promin})\left(  \frac{y_j-17.5}{20-17.5}\right), & 17.5 \leqslant y_j < 20 \\
T_{promax} + \frac{1}{2}\left( T_{co} - T_{promax}\right)\left( \tanh{\frac{y_j - 20}{0.1}+1}\right), & y_j \geqslant 20,
\end{cases}
\label{eq:initial temperature}
\end{align}
where the $y$ values are in Mm, $T_{ch}$ is the temperature of the chromosphere = $(8 \times 10^3)$, $T_{co}$ of the corona = $(1.8 \times 10^6)$, $T_{promin}$ is the minimum temperature of the prominence= = $(6 \times 10^3)$ and $T_{promax}$ is the maximum temperature of the prominence = $(1.4 \times 10^4)$. This sets a temperature of $8000$~K in the chromosphere, transitioning to 1.8 $\times 10^6$~K for the corona, dropping at $y=y_b$ to $6000$~K for the lower part of prominence and an increasing linear profile from $6000$ to $14000$~K for the upper part of the prominence ending at 20~Mm. The different temperature regions in the vertical direction of the stratified solar atmosphere are connected smoothly using a hyperbolic tangent function. Then, the corresponding gas pressure and density stratifications are derived, ensuring magnetohydrostatic balance as expressed in Eq.~(\ref{eq:mhsBalance}).  Under the assumption of a fully ionized plasma with a 10:1 abundance ratio of H: He\vertwo{, $\left(\frac{n_H}{n_{H_e}} = 10\right)$,} we have the relation $\rho = 1.4 m_p n_H$, where $m_p$ is the proton mass and $n_H$ the number density of hydrogen. Up to heights $y=y_b$, we simply compute the 1D pressure and density array at an arbitrarily high resolution from the ideal gas law combined with the discrete formula,
\begin{equation}
p_j = \frac{p_{j-1} + \frac{\Delta y (g_j + g_{j-1})\rho_{j-1}}{4}}{1-\frac{\Delta y(g_j + g_{j-1})}{4 T_j}},
\label{pressure_init}
\end{equation}
where $g_j$ is the local solar gravity value given as $g(y) = -2.74 \times 10^4(R_\odot^2/ (R_\odot + y)^2)$~cm~s$^{-2}$ and $j$ is a grid index along the vertical direction. For $y>y_b$, a similar formula is used where 
the Lorentz force is additionally accounted for. The $B_z$ field, in this case, is defined by the following formula,
\begin{equation}
B_{z_{j}} = \left(B_{z,j} \cdot e^{ -\frac{(y_{j-1}-y_b)}{\lambda_B} }\right)^2 + \left(B_{z,j} \cdot e^{-\frac{y_j -y_b}{\lambda_B}}\right)^2 ,
\end{equation}
and we solve the corresponding pressure as in Eq.~\ref{pressure_init}, which is given as follows,

\begin{equation}
 p_j = \frac{\left(p_{j-1} + \frac{\Delta y (g_j + g_{j-1})\rho_{j-1}}{4} + \frac{\Delta y}{\lambda_B}\cdot \frac{B_{z_{j}}}{2.0} \right)}{{1-\frac{\Delta y(g_j + g_{j-1})}{4 T_j}}}.   
\end{equation}

The self-consistently derived density profile corresponding to the prescribed temperature stratification is then shown in Figure~\ref{init_temp}. The simulation parameters thus represent conditions typical to a quiescent prominence. From a modeling perspective, the most relevant aspect of capturing prominence dynamics is to adopt the right density and temperature contrast realized in prominences while allowing for a fully gravitationally stratified atmosphere in a magnetized environment. Our initial conditions match these contrasts and adopt a temperature stratification (in line with previous model efforts as in \citet{2014ApJ...789...22K}).  \vertwo{The two-order of magnitude density inversion present at the prominence lower edge matches that extensively studied in earlier local box studies} (\citet[][]{hillier2012numerical, hillier2012numerical2,keppens2015solar,xia2016formation,2016ApJ...820..125T}). The adopted magnetic field is weak (4G), and at the lower edge of the quiescent prominence range, but we note that the pressure is derived from the magnetohydrostatic equilibrium. The more important dimensionless parameter called plasma~$\beta$ is shown as a function of height at different times in our later Figure~\ref{plasma_beta}. Most of the time, but especially during the initial stages, we remain below unity throughout the corona. We note that values up to plasma beta unity and above are reported in various works on prominences (see the review in \citet{2018LRSP...15....7G}).  Our $t=0$ model ingredients are in accord with current understandings from ideal MHD modeling.

The boundary conditions are set to be periodic on the left and right sides, and the top and bottom boundary conditions were specially handled in the simulation setup. The boundary values at the bottom were fixed for all time steps; for the top boundary, we extrapolate the density and pressure from the inner cells assuming hydrostatic equilibrium. In contrast, the magnetic field is extrapolated assuming zero normal gradients, and the velocity values are copied from the inner cells into the ghost cells.

The region of higher density in the corona mimics our prominence body, extending from the lower prominence corona transition region at $y_b=11.25$~Mm up to a height of 20~Mm. We initiate the RT instability by inducing a multimode perturbation within the lower prominence-corona interface. This perturbation is a superposition of 50 small-amplitude sinusoidal velocity fields, periodic in the $x$-direction with random phases. As the initial condition is unstable, the perturbation will deform the interface to form bubbles and pillars. The resulting mixing in the presence of the mean magnetic field is dependent on the angle between the perturbation wave vector $\textbf{k}$ and the mean magnetic field $\textbf{B}$, where the tension of the field can suppress the growth of the instability. This effect was first studied analytically by \cite{chandrasekhar1961international}. The dispersion relation in a simplified two-layer, plane-parallel model in the presence of a magnetic field \vertwo{under the assumption of incompressibility} is given by
\begin{equation}
	\sigma^2 = gA \textit{k} - \frac{(\textbf{k}\cdot {\bf B})^2}{2\pi (\rho_{+} + \rho_{-})},\label{eq:rtiGrowth}
\end{equation}
where $\sigma$ is the growth rate of the instability, $A$ is the Atwood number given by $A = \frac{\rho_{+}-\rho_{-}}{\rho_{+}+\rho_{-}}$, and where we use cgs units. $\rho_{+}$ is the density of the heavier fluid, and $\rho_{-}$ is the density of the lighter fluid. The aforementioned perturbation is in the form of an incompressible flow that initializes both the horizontal and vertical velocity components localized near the $y_b$ interface. The spatial amplitude of the perturbation is set to $50$~km. The critical wavelength $\lambda_c$ below which the magnetic RT instability is suppressed for a chosen magnetic field strength and angle is given by,
\begin{equation}
\lambda_c = \frac{B^2 \cos^2 \vartheta}{g (\rho_{+} - \rho_{-})}, \label{eq:waveCrit}
\end{equation}
where $\vartheta$ is the angle between the wavevector $\textbf{k}$ and the field $\textbf{B}$. This angle is important in studying the RT instability where the magnetic field shear affects the growth rate of RT instability, which is bounded in its presence \citep{2014ApJ...785..110R}. As our setup has the magnetic field almost perpendicular to the simulated plane (which contains $\textbf{k}=k_x\textbf{e}_x$, making $\theta=\vartheta$), the stabilizing effect of magnetic tension is minimal as $\sigma^2 \rightarrow gAk$. This implies instability despite the otherwise stabilizing effects of magnetic fields. At exact perpendicular orientations ($\vartheta=90 \degree$), the growth of the RT instability is most susceptible to perturbations at the shortest wavelengths since $\lambda_c \rightarrow 0$. We note that a true 3D setup will allow for even more freedom, but our chosen angle for the magnetic field (nearly orthogonal to the plane shown) is such that our setup is close to realizing perfect $\textbf{k}\cdot\textbf{B}=0$ conditions. This well-known minimal field line bending requirement determines the stability of low beta plasmas in the laboratory and astrophysical settings (\citet{goedbloed2019magnetohydrodynamics}). Finally, the Atwood number can range between 0 and 1, where the density contrast between the fluids (here, $A\approx1$) is directly proportional to the growth rate of the instability \citep{2021PhyD..42332838Z}. Note that the presence of magnetic shear, as present in our setup at the bottom prominence boundary, in addition to solving for compressible evolutions, makes the simple two-layer, incompressible predictions of Eqns.~(\ref{eq:rtiGrowth}) and (\ref{eq:waveCrit}) only valid approximately.

\vspace{0.5cm}
\subsection{Overall evolution to fully developed turbulence}

\begin{figure*}[htbp!]
\centering

   \includegraphics[width=18.5cm]{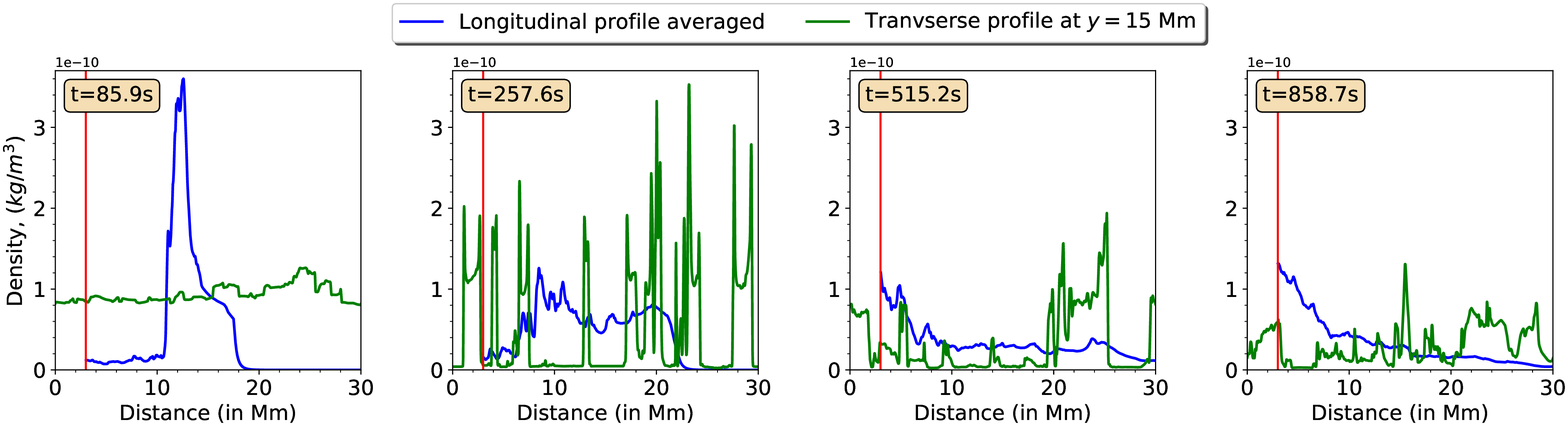}
   \includegraphics[width=18.5cm]{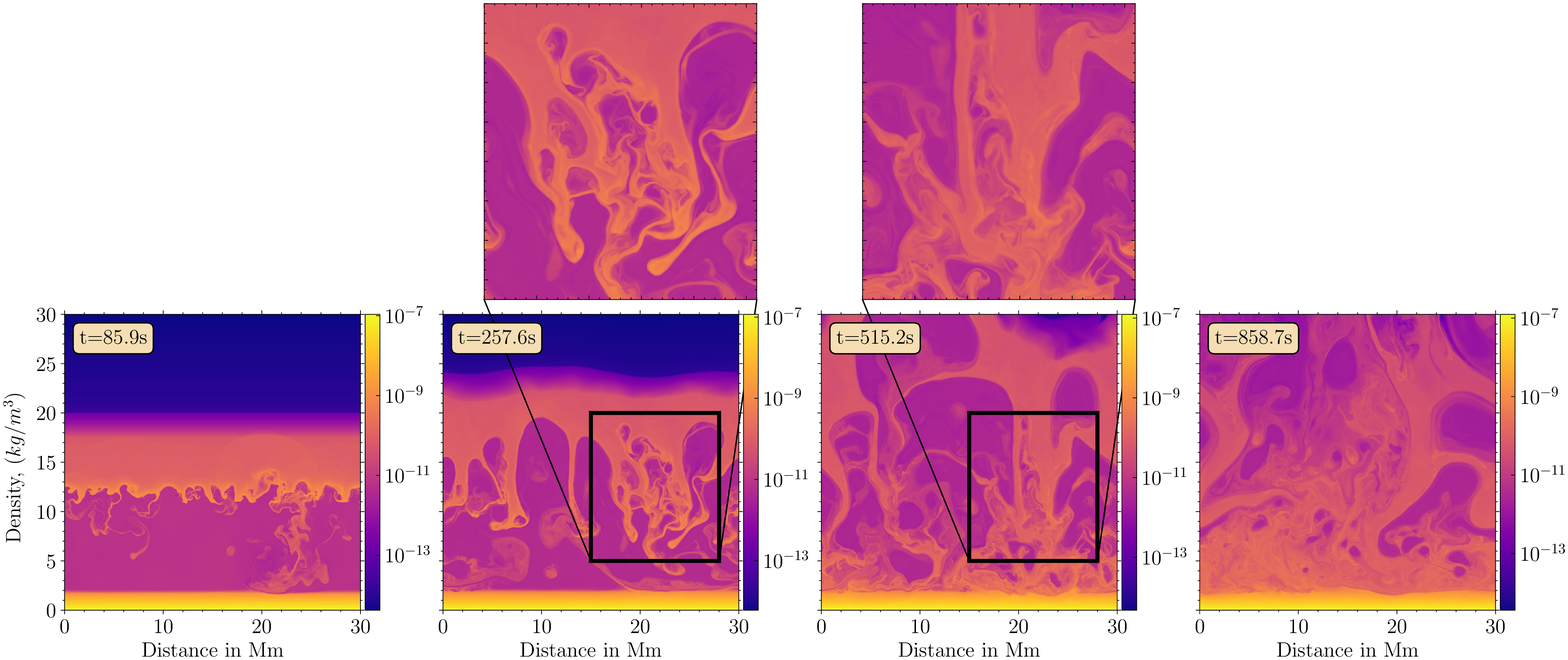}
   \includegraphics[width=18.5cm]{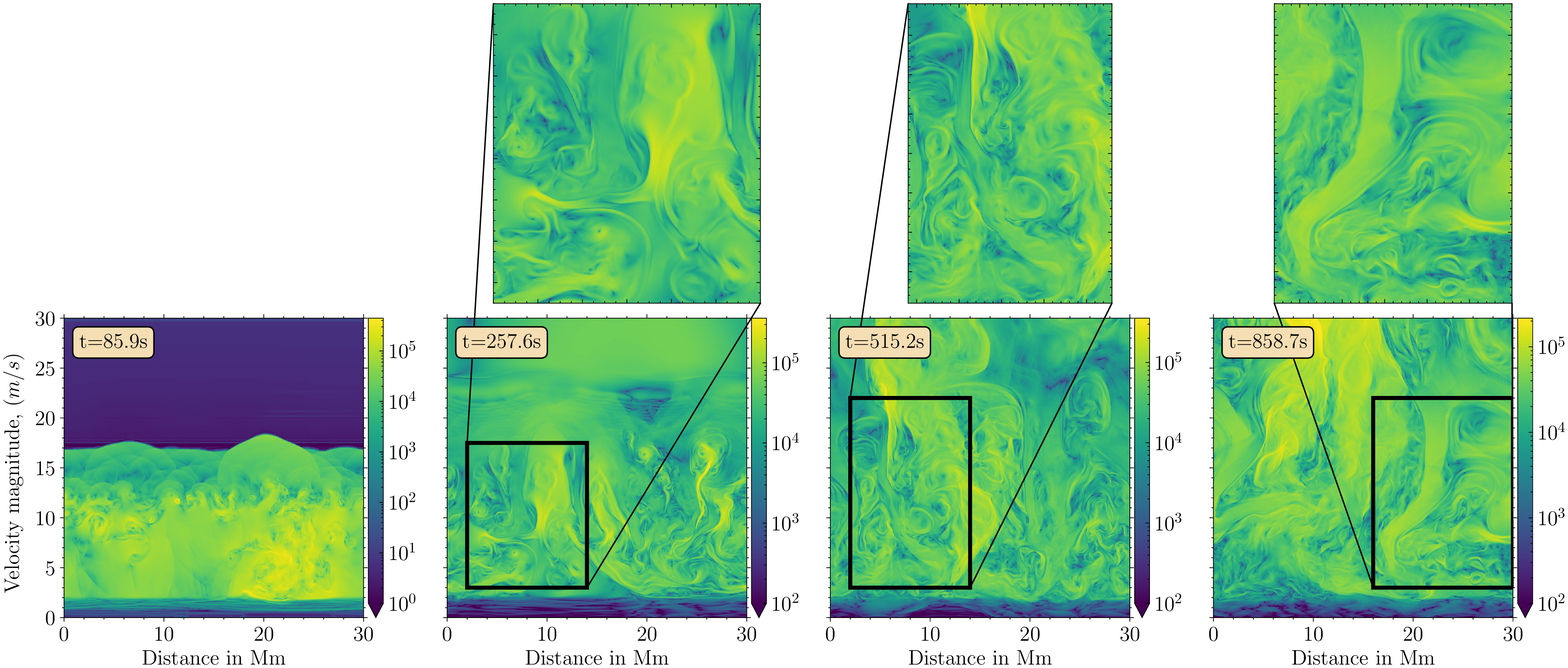}
     \caption{\textbf{Density profile (top) for longitudinally averaged (blue) and transverse profile at height $y=15$~Mm (green) for each time.} Time evolution of the density (middle) and velocity magnitude (bottom) for a case of $\theta = 88.5\degree$. At time 85.9~s, the RT instability forms at the lower prominence-corona interface, corresponding to the early nonlinear phase. At time 257.6~s, the perturbed layer undergoes deformation due to the RT instability and forms bubbles and plumes as it transits to the later non-linear phase. At time 515.2~s, the non-linear phase further develops as part of the matter gets reflected upwards, and we see the formation of further substructures on the edges of bubbles and plumes. At time 858.7~s, many small-scale structures in the prominence formed, which transfer energy to intermediate scales, and turbulence is prominent at this time.}
     \label{vel_evolution}
\end{figure*}

\begin{figure}[htbp!]
\centering
    \resizebox{\hsize}{!}{\includegraphics{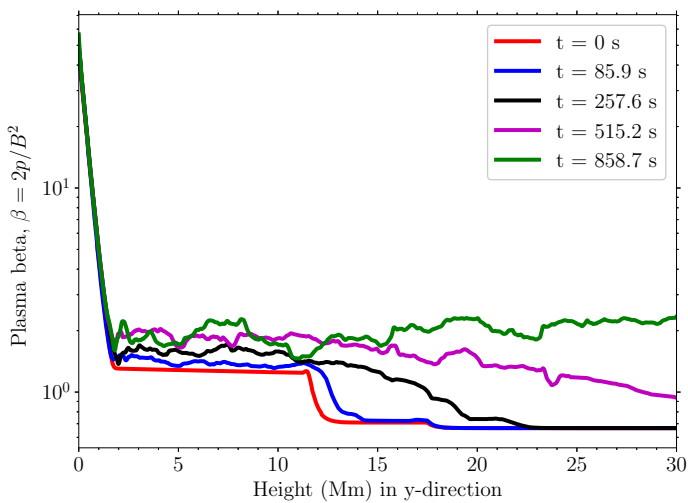}}
    \caption{Local plasma beta (averaged in the $x$-direction) in the initial prominence of the reference case $\theta = 88.5\degree$ at $t = 0$~s (red) compared to its evolution in time at $t =$ 85.9~s, 257.6~s, 515.2~s, and at t = 858.7~s.}
	\label{plasma_beta}
\end{figure}

\begin{figure*}[htbp!]
\centering
    \includegraphics[width=18.5cm]{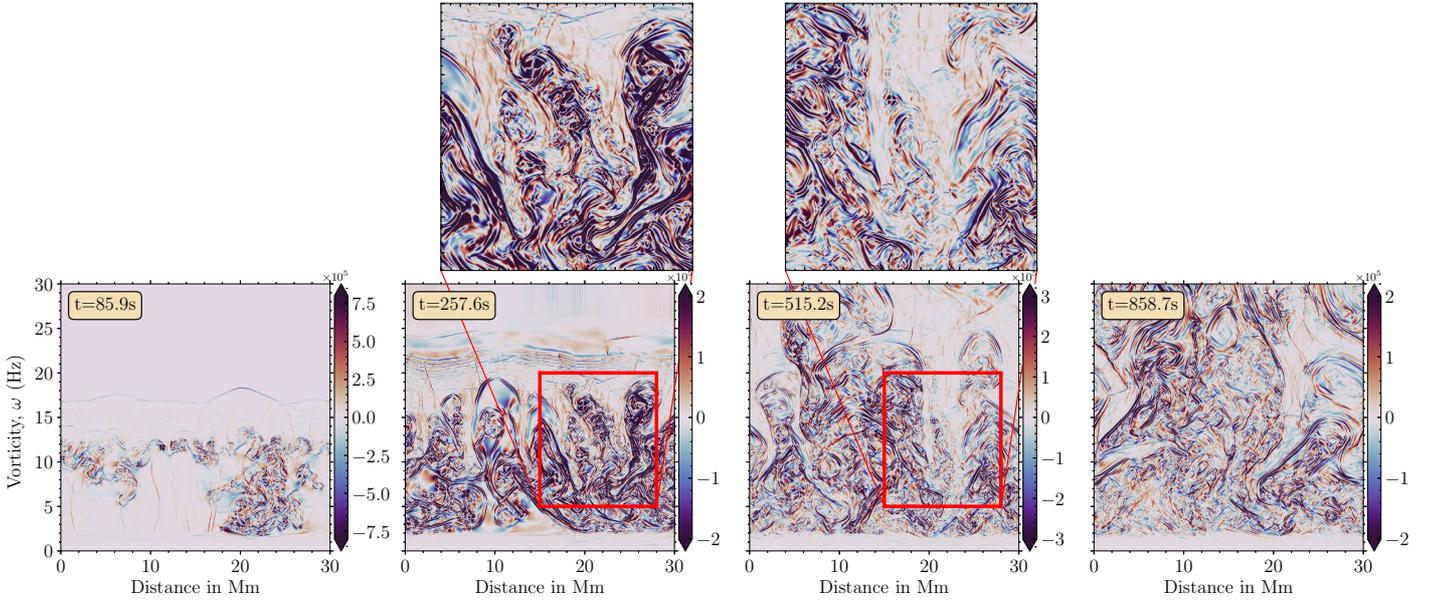}
    \caption{Time evolution of the vorticity field for a case of $\theta=88.5\degree$. The initial stages show the onset of RT instability, followed by the formation of vortical structures in the shape of bubbles and plumes at 257.6~s. This then transitions to the late non-linear phase at time 515.2~s, with the formation of small-scale structures due to the increasing shear at the edge of bubbles and plumes. At time 858.7~s, these sub-structures organize in vortices and other coherent structures, predominantly in the vertical direction.}
    \label{vort_evolution}
\end{figure*}

\begin{figure*}[htbp!]
\centering
   \includegraphics[width=18.5cm]{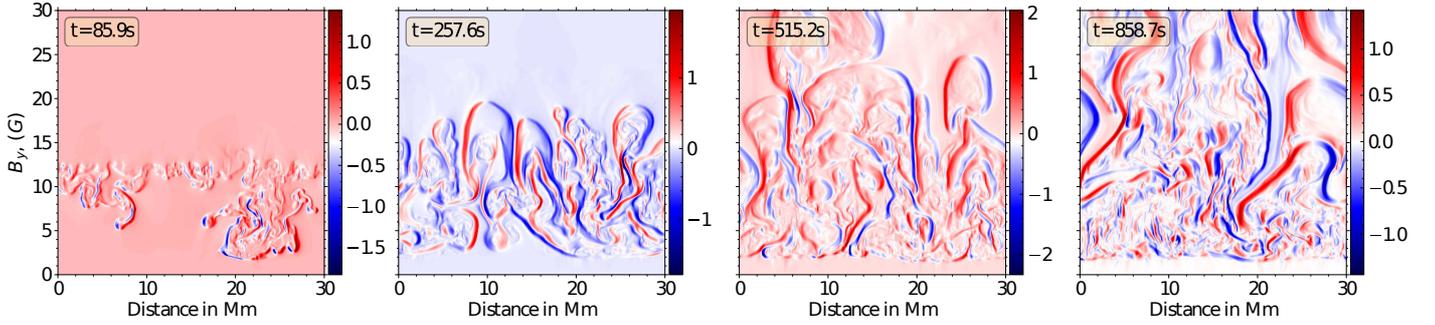}
     \caption{Time evolution of $B_y$ (magnetic field component in the stratified $y$-direction) for a case of $\theta =88.5\degree$. The presence of many locally anti-parallel poloidal magnetic field lines is seen as the prominence flow transitions from a linear to a non-linear phase.}
     \label{by_evolution}
\end{figure*}

\begin{figure*}[htbp!]
\centering
   \includegraphics[width=17.5cm]{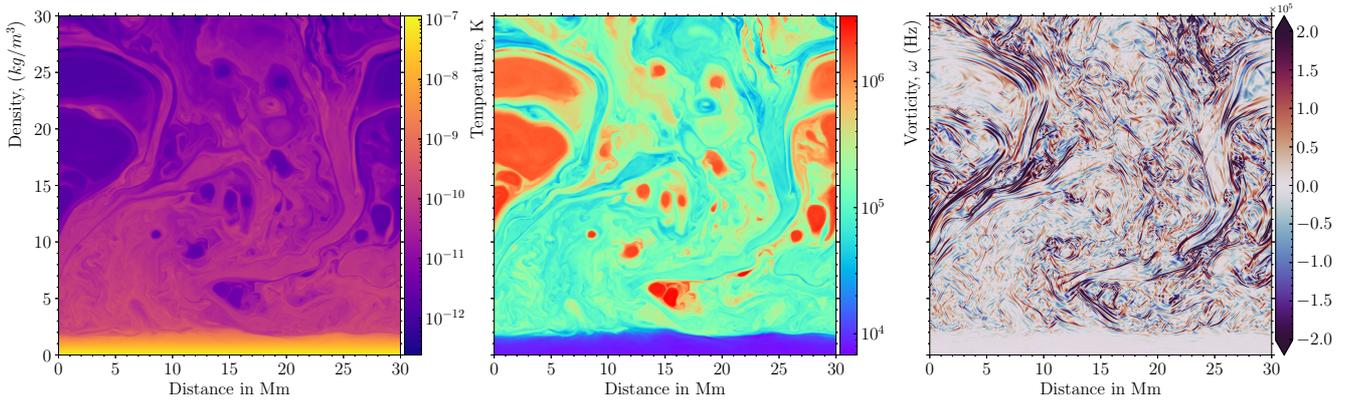}
     \caption{Late time evolution of density, temperature, and vorticity field for a case of $\theta =88.5\degree$ at time=1073.38~s \vertwo{when the prominence has dissipated.}}
     \label{latetime_evolution}
\end{figure*}

\begin{figure}[htbp!]
	\centering
	\includegraphics[width=8cm]{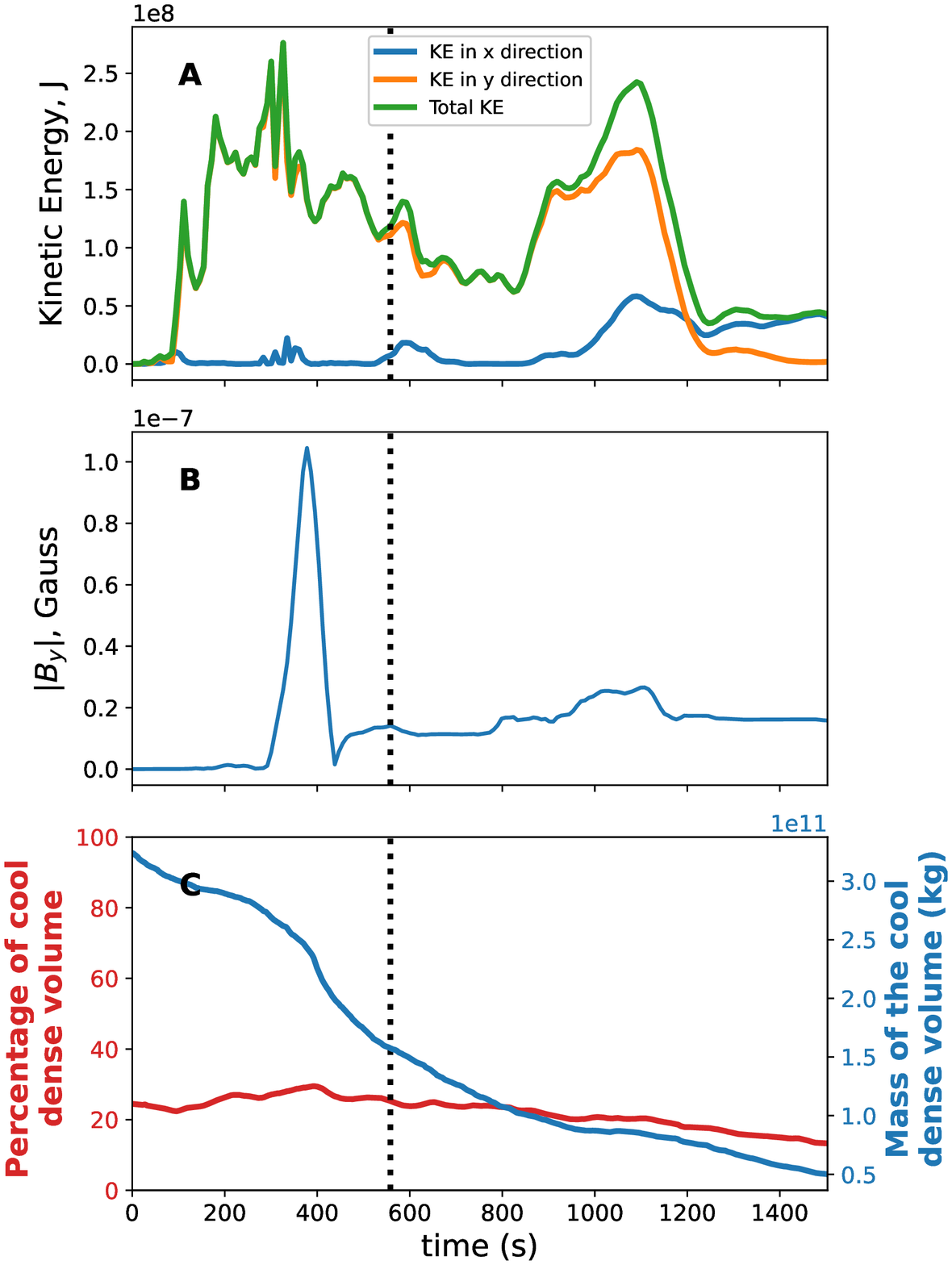}
	\caption{Panel \textbf{A}: The time evolution of the volume-averaged kinetic energy across the domain in $x$ and $y$ directions and volume-averaged total kinetic energy of the prominence. Panel \textbf{B}: The time evolution of the mean magnitude of the prominence's vertical magnetic component $B_y$ (volume-averaged). Panel \textbf{C}: The evolution over time of the volume percentage of cool dense solar prominence (+ coronal interface) material, that is, excluding the chromosphere, taken below a threshold temperature of \vertwo{$5 \times 10^4$~K}, and denser than $10^{-11}$~kg~m$^{-3}$ (cf. Figure~\ref{init_temp}), is plotted together with its corresponding mass. \vertwo{The black dotted line in Panel \textbf{C} represents the time 558.15~s, when the turbulence analysis is done.}}
	\label{time evol ke by}
\end{figure}

Figure~\ref{vel_evolution} presents the time evolution of density and velocity in our ideal MHD simulation of the RT instability within a quiescent prominence for $\theta = 88.5 \degree$. From Figure~\ref{vel_evolution}, we see how the density and velocity field magnitude evolves for this case. From an initial multi-mode perturbation, the instability sets in due to the effect of gravity, and vertical structures begin to form. These coherent structures in the form of pillars and bubbles gradually show more and more fine structures. At time 85.9~s, when we have entered our early non-linear stage, we see falling fingers or pillars and rising mushroom-shaped structures. From the dispersion relation \citep{chandrasekhar1961international,1982GAM....21.....P}, the growth of small-scale modes is fast compared to the large scales. The presence of non-vanishing poloidal (i.e., $B_{XY}$ plane) magnetic field components produces tension forces on the small scales, which reduces shear and mixing between the fluids and influences the displacement rate of bubbles and fingers from the interface.

The combination of an increased gas pressure against a decreasing magnetic pressure results in $\beta$ increasing in time with the growing level of turbulence. As the simulation starts from an initial state of a quiescent prominence condition having a volume-averaged unit plasma $\beta$, the influence of magnetic pressure is important during the linear phase. As the non-linear phase develops, the plasma $\beta$ value increases throughout the transition of the system to a turbulent state, where the plasma pressure also affects the energy cascading processes through the formation of small-scale structures, as can be seen from Figure~\ref{plasma_beta}. During the turbulent phase (see $t=858.7$~s), the plasma $\beta$ value increases considerably above the top of the initial prominence, where the plasma pressure becomes more dominant as compared to the magnetic pressure. The turbulence affects the detailed balance between kinetic, thermal, and magnetic energy. The local vertical profile of plasma beta at selected times is just one way of quantifying this transition. The buoyant rising of bubble-like structure seen in Figure~\ref{vel_evolution}, in conjunction with the falling dense fingers, give instantaneous density variations at a height $y =$~15~Mm as shown in Figure~\ref{vel_evolution}, top panels (green lines). These horizontal cuts show how density contrasts indeed remain strong between the pillars and plumes. Some matter is also entrained from the upper chromospheric regions on impacting the transition region, and then gets entrained up into our coronal volume. At late stages, the higher regions become beta unity and above, with buoyant matter rising through our top boundary and fine-scale mixing, causing the thermal pressure to rise.

The various stages and the gradual trend to more and more fine-scale turbulence, especially after the falling pillars partially reflected off transition region heights and mix up and downward moving plasma, is particularly evident with the plots of vorticity in Figure~\ref{vort_evolution}.

The magnetic field is initially set up to exert an upward magnetic pressure force to lift the prominence material locally. An evolving $B_y(x,y)$ magnetic field component is established during the simulation. This is shown in Figure~\ref{by_evolution}, and one can note the development of many locations of nearby anti-parallel poloidal (in-plane) field lines. This naturally plays a role in the (numerical) dissipative exchange of energy from magnetic to kinetic. After 257~s, one can even see shocks forming on top of the prominence. These shocks develop in the original, highly evacuated layer above the prominence, i.e., where we have the derived initial density very low and almost vacuum in nature. The shocks ultimately deform this top layer before it disappears through our open top boundary in the transient phase before about 500~s. 

Figure~\ref{latetime_evolution} describes the density, temperature, and vorticity field of the truly turbulent state that is found at a later time. Indeed, the cool, dense material of the prominence collides with the transition region marking rapid changes in density and temperature, which leads to a significant rise of kinetic energy in the horizontal direction, as seen from Figure~\ref{time evol ke by}(A). After 800~s, we consider the prominence to transition to a fully turbulent state. In these nonlinear, turbulent stages, the full coronal region is to be considered as our prominence structure, which is now redistributed throughout the coronal domain. This is also seen in the way the $x$-averaged density profile above the transition region, as indicated in the blue lines in the top panels of Figure~\ref{vel_evolution}, shows how most of the original prominence mass (from the integral under this curve) remains in the entire coronal region.

Figure~\ref{time evol ke by}A shows the evolution of the volume-averaged kinetic energy across the domain. The kinetic energy in the $y$-direction increases during the transient period of shock development in the region above the initial prominence. Furthermore, the sharp increase of the magnetic field component in the $y$-direction is also seen to succeed in this increase in kinetic energy according to Figure~\ref{time evol ke by}B (see also Figure~\ref{plasma_beta}). At these times, the falling material gets reflected back up from the sharp transition region gradients. Hence, the increase in the vertical kinetic energy describes the formation of the falling fingers and rising plumes associated with the linear and non-linear phases of the RT instability. The rise in the height of the upper prominence boundary is mainly driven by the development of the bubbles, specifically. For the general RT instability, cool dense material gets accelerated in the downward direction in the presence of gravity. With time, the mixing of the plasma is then responsible for an increase in the volume of the cool dense material. In Figure~\ref{time evol ke by}C, the time evolution of the percentage of the volume of cool material is plotted together with its mass. This is calculated by considering only the plasma below an upper threshold temperature of $5 \times 10^4$~K and above a threshold density of $10^{-11}$~kg~m$^{-3}$. We can see that overall there is a slight decrease in the volume with a more significant decrease in the corresponding mass through the turbulent stages. This suggests the degree of cool material that transits through the top boundary is partially compensated by the mixing within the domain above the chromosphere boundary.

From Figure~\ref{vel_evolution} \&~\ref{vort_evolution} at time 515.2~s, we see the development of substructures along the edges of the large-scale bubbles and plumes, which are shown by the zoomed insets. \citet{keppens2015solar} discussed the appearance of these small-scale structures in their simulation due to the strong shear flows established along the edges of the bubbles. These authors simulated for only seven minutes in physical time and, as such, did not (yet) achieve the fully turbulent stages that we will go on to analyze in more detail. These substructures lead to the formation of the KH instability and the vortices along the vertical direction, as seen from the velocity magnitude plots marked by the squares in Figure~\ref{vel_evolution}. \citet{2019ApJ...874...57M} discussed evidence of magnetic RT unstable plumes in a loop-like eruptive prominence. They suggested that the nonlinear phase of this magnetic RT-unstable plume collapses due to the formation of a KH vortex in the downfalling plasma. We observe the same process of downfalling prominence material accompanied by KH vortices in the non-linear phase of our study. %They suggested that the vertical and horizontal threads are really different views on the same RT-related structures of prominence.
%\citet{2016ApJ...825L..29X} %performed 3D MHD simulations on the internal dynamics of a twin-layer prominence which \je{developed the falling fingers and uprising bubbles characteristic of the RT instability}. During the 1000~s run of their simulation, they showed that \je{during} the turbulent phase, there was a strong coherent evolution between both layers, \je{attributed} to the magnetic connectivity.}

In our study, the evolution of density structures shows turbulence developing in the prominence, indicated by the fine-scale structures present in Figure~\ref{vel_evolution}. During time 858.7~s, the kinetic energy evolution from Figure~\ref{time evol ke by}A, shows interesting behavior. The formation of small-scale structures in the horizontal, i.e., $x$-direction, coincides with the increase in kinetic energy in the same direction. In the non-linear stage, the cool and dense material which is accelerated downwards due to gravity, hits the evolving transition region and collects matter and energy from the chromosphere. This leads to an increase in fluctuations and build-up of kinetic energy in the horizontal direction. Until then, the kinetic energy was predominantly vertical, attributed to the linear and non-linear stage development of the RT instability. Thus, the formation of vortices and swirling structures transfer energy among the scales and in both directions in this later stage. This turbulent stage is then analyzed in what follows.

\section{Results for the turbulent stages}\label{sec-ana}

\subsection{Observational counterparts}

Quiescent prominences display a wide range of dynamics and structures predominantly oriented in the vertical direction \citep{Berger_2008}. The vertical prominence structures are commonly characterized by dark upflows and the constant motion of streaming downflows, vortices, transverse oscillations, and expanding arch events. On the other hand, the horizontal prominence structure shows relatively little motion, with dynamics seemingly restricted to the appearance of short transient flows. The observational and theoretical work of \citet{leonardis2012turbulent} suggests these features in quiescent prominence are together suggestive of turbulence. They applied statistical methods to the images from Hinode Solar Optical Telescope Ca $\RN{2}$ H-Line observations to quantify the turbulent characteristics of a long-lived quiescent prominence. They investigated the multifractal and intermittent statistical scaling of turbulence properties for a long-lived quiescent prominence. Such turbulence in a finite region showed non-Gaussian statistics, multifractality, and extended self-similarity. These statistics gave information on scale behavior and correlation amongst scales and revealed non-Gaussianity through power law distributions for the intensity fluctuations measured in time and space. 

Similarly, \citet{freed2016analysis} used Ca $\RN{2}$ and $H \alpha$ observations from Hinode to investigate the kinetic energy and vorticity associated with plasma flows in a quiescent prominence and found indices to the power law fit to be in the range of $-1$ to $-1.6$. \citet{hillier2017investigating} used $H \alpha$ Dopplergrams from Hinode SOT observations to investigate the turbulent nature of a prominence using structure functions and showed the existence of two distinct regions with a break in the power law at around $2000$~km. They also determined the non-Gaussianity and intermittent dynamics of the turbulence found within the observed prominence. The authors concluded that the heating rate due to turbulent dissipation is very small and unimportant in heating the prominence plasma. However, the diffusion due to magnetic reconnection is of a similar order to the estimated ambipolar diffusion and a few orders of magnitude greater than the Ohmic diffusion.
These turbulent characteristics of plasma flow within solar prominences have so far primarily been addressed using observations. These studies focused on vertical and horizontal dynamics, where the former is characterized by extended, long-lived striations, and the latter form more short-lived, transient phenomena. The observational works analyze turbulence in terms of non-Gaussian statistics, multifractality, and extended self-similarity. In the following paragraphs, we will go into more detail about each of these properties. The main aim of this work is to demonstrate that our simulated prominence exhibits properties in its turbulent stage that are very similar to the observed counterparts.

\subsection{Analysis of the simulations}

\begin{figure}[htbp!]
\centering
	\resizebox{\hsize}{!}{\includegraphics{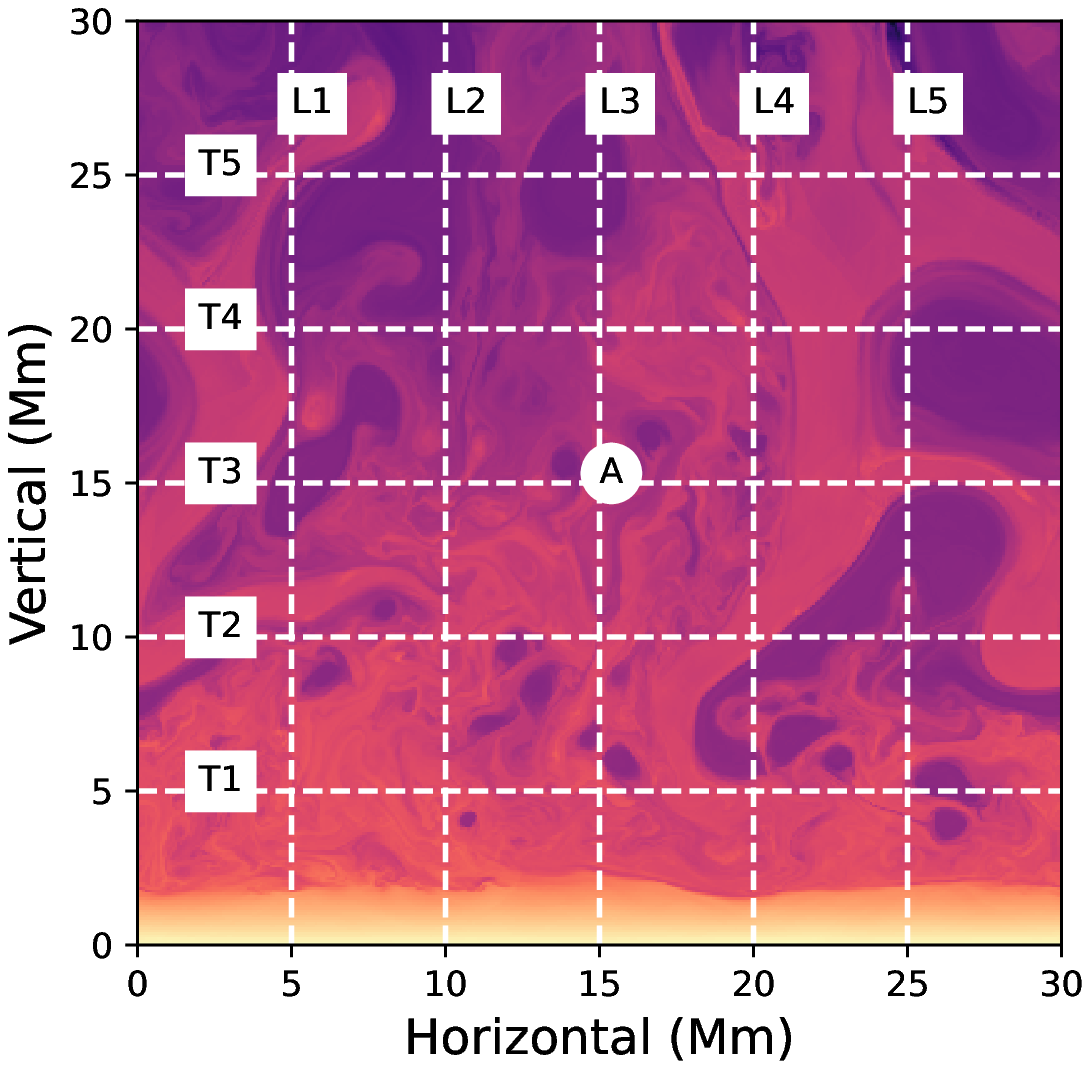}}
	\caption{The data has been divided into five longitudinal strips, L1-L5, and five transverse strips T1-T5, where the spatial analysis has been done for the fields at specific times. We also take time-sampled data at point A, where the temporal evolution of the data is stored at a cadence of 0.858~s (the density field of the solar prominence at time 858.7~s is shown in the background).}
	\label{setup}
\end{figure}

In order to understand the characteristics of the turbulence within our simulation, we process our MHD data and apply statistical methods. We use data at the full spatial resolution $1280 \times 1280$ at specific times. We also isolate a subset of the data spanning $\sim 25$~minutes, with a cadence of $\sim 0.858$~s, that covers the period of RT instability evolution within the prominence. Similar to the work of \citet{leonardis2012turbulent}, we use five strips each in the longitudinal and transverse directions, the former corresponding to the dominating vertical motion of flow in the prominence as seen in Figure~\ref{setup}. For analysis, we consider the domain above the upper chromosphere boundary. Each longitudinal (transverse) strip has 1131 (1280) grid points along it, so a resolution of approximately 23~km. In practice, we take small ensembles consisting of $11$ (five on either side of the center, inclusive) neighboring rows (columns) of grid cells for each strip in the transverse (longitudinal) direction. The statistical quantities are calculated for each of the 11 strips before averaging across this strip width to have more consistent statistics and capture the variation of any field in the spatial domain. Additionally, we use a single central point A within the domain, as shown in Figure~\ref{setup}, to obtain the high-resolution time series of the temporal fluctuations within the simulation using the particle module of {\tt MPI-AMRVAC}. We apply all our statistics to the fluctuating values of the field variables (velocity, magnetic components, or magnitude). We also apply statistics to the temperature field to enable a meaningful comparison to previous observational results. These fluctuating values of the field variables are given by $I'(u)$, which is obtained by subtracting the mean component from the instantaneous components of the fields, i.e., $I'(u) =I(u) - \langle I(u) \rangle$, where $u$ is either a spatial coordinate $x$, or the temporal state $t$, and the mean is $\langle I(u) \rangle=\int_{u_{min}}^{u_{max}} I(u) \, du/(u_{max}-u_{min})$ in the range $u\in[u_{min},u_{max}]$. In what follows, we will simplify the notation and hence write, e.g., $V(r)\equiv V'(r)$ for the fluctuation in the velocity field at location $r$ instead.

\begin{figure*}[htbp!]
\centering
    \includegraphics[width=12cm]{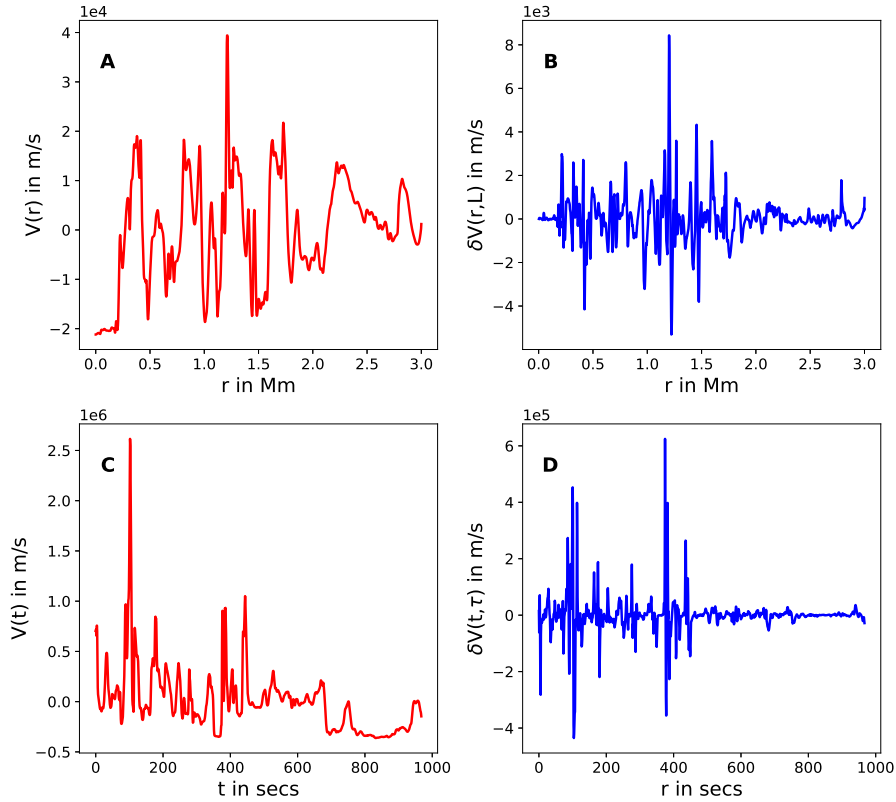}
    \caption{Spatial series of the velocity field \textit{V(r)} for strip L5 (A) and the corresponding first difference, $\delta V = V(r+L) - V(r)$ with $L=1$ lag $\sim 23$ km (B) for the prominence at time 858.7~s. The temporal series of the velocity field \textit{V(t)} from time 747.01~s to 1747.01~s, for point A in Figure~\ref{setup} is shown in (C). Its corresponding first difference $\delta V = V(t+ \tau) - V(t)$ with $\tau = 1$ lag $\sim 0.858$~s is shown in (D).} 
    \label{spatial_first_diff_series}
\end{figure*}

For the spatial analysis, we analyze the two representative snapshots at time $558.15$~s and $858.7$~s in our simulation, where the structures in the transverse, i.e., $x$-direction, start to grow compared to the longitudinal direction as seen in Figure~\ref{time evol ke by}A, and we observe a turbulent state in our simulation. This can also be seen from the snapshots shown in Figure~\ref{vel_evolution}, where the mixing has led to the formation of large-scale structures in both directions, deforming the clear bubble and plume shapes of the RT instability. In the statistical analysis, we take differences or increments at varying spatial and temporal separations (lags) denoted by $L$ and $\tau$, respectively. In Figure~\ref{spatial_first_diff_series}, we show the instantaneous (strip-width averaged) spatial variation in the velocity field for the strip $L5$. In panel A we show the raw $V(r)$, and its corresponding velocity increment of first differences $ \delta V = V(r + L) - V(r)$ in panel B, where $r$ is the spatial coordinate (where $r$ is either $x$ or $y$ for horizontal/vertical strips). For a spatial lag of $L = 23$~km along the strip, which hence subtracts the instantaneous neighbor value, panel B presents the spatial velocity series in the first difference $\delta V(r,L=1)$, which we find to be highly fluctuating with a consistent standard deviation across the mean. In an analogous fashion, the temporal series for the sample point A in the setup, indicated in Figure~\ref{setup} from time 747.01~s to 1747.01~s, is shown in Figure~\ref{spatial_first_diff_series} (in panels C and D), for the temporal variation of the velocity field i.e. $V(t)$ (panel C) and its corresponding time increment of first difference $\delta V (t,\tau) = V(t+ \tau) - V(t)$ (panel D) where $t$ is the temporal scale and the temporal lag $\tau = 0.858$~s.

\vspace{0.5cm}
\subsection{Correlation time scales}
\begin{figure}[htbp!]
\centering
	\resizebox{\hsize}{!}{\includegraphics{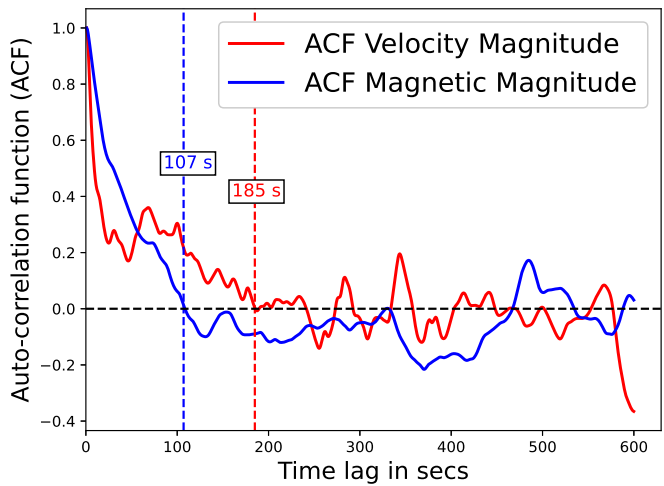}}
    \caption{Normalized temporal autocorrelation function ($R(\tau)$) for the velocity and magnetic fields at the center of the simulation box (Point A in Figure~\ref{setup}). The correlation time seen as the first zero crossing gives the correlation timescale of 185~s for the velocity field and the correlation timescale of 107~s for the magnetic field.}
    \label{acf_time}
\end{figure}

We analyze the correlation time for the velocity and magnetic fields using the autocorrelation function normalized by the variance, i.e., the correlation coefficient defined as,
\begin{equation}
R(\tau) = \frac{\langle I(t) I(t + \tau)\rangle}{\langle I(t)^2\rangle},
\label{acf equation}
\end{equation}
where $\langle .\rangle$ denotes the average over the temporal direction $t$. This autocorrelation function is thus a function of the lag $\tau$, which we vary in what follows in the range of $0.85$\,--$\,600$~s for our analysis. The correlation coefficient value lies between -1 (perfectly anticorrelated), and +1 (perfectly correlated), while a zero value indicates uncorrelated data. Since turbulence is a random and stochastic phenomenon, the autocorrelation function converges to zero for the limit of a very large time lag, and one defines the correlation time scale $T_c$ of the turbulence as,
\begin{equation}
T_c = \int_{0}^{\infty}R(\tau) d \tau \,.
\end{equation} 
Due to turbulence, the temporal correlation of the processes diminishes as the lag time $\tau$ increases. Due to this, the integral converges to give the correlation (or integral) timescale. The value of $T_c$ gives the measure of memory of the turbulence \citep{pope2001turbulent}. In the case of a finite time sample series, one frequently uses an approximate $T_0$, corresponding to the first zero crossing of the function $R(\tau=T_0)=0$, to quantify this correlation time. From Figure~\ref{acf_time}, we see that in our simulation data, we have a correlation time $T_0$ of around 185~s for the velocity field and around 107~s for the magnetic field. A comparison can be made with \citet{hillier2017investigating} who found the correlation time for velocity to be 328~s. The intensity was 544~s, calculated using the half-width half-maximum (HWHM) of the auto-correlation. 

\begin{figure}[htbp!]
\centering
    \resizebox{\hsize}{!}{\includegraphics{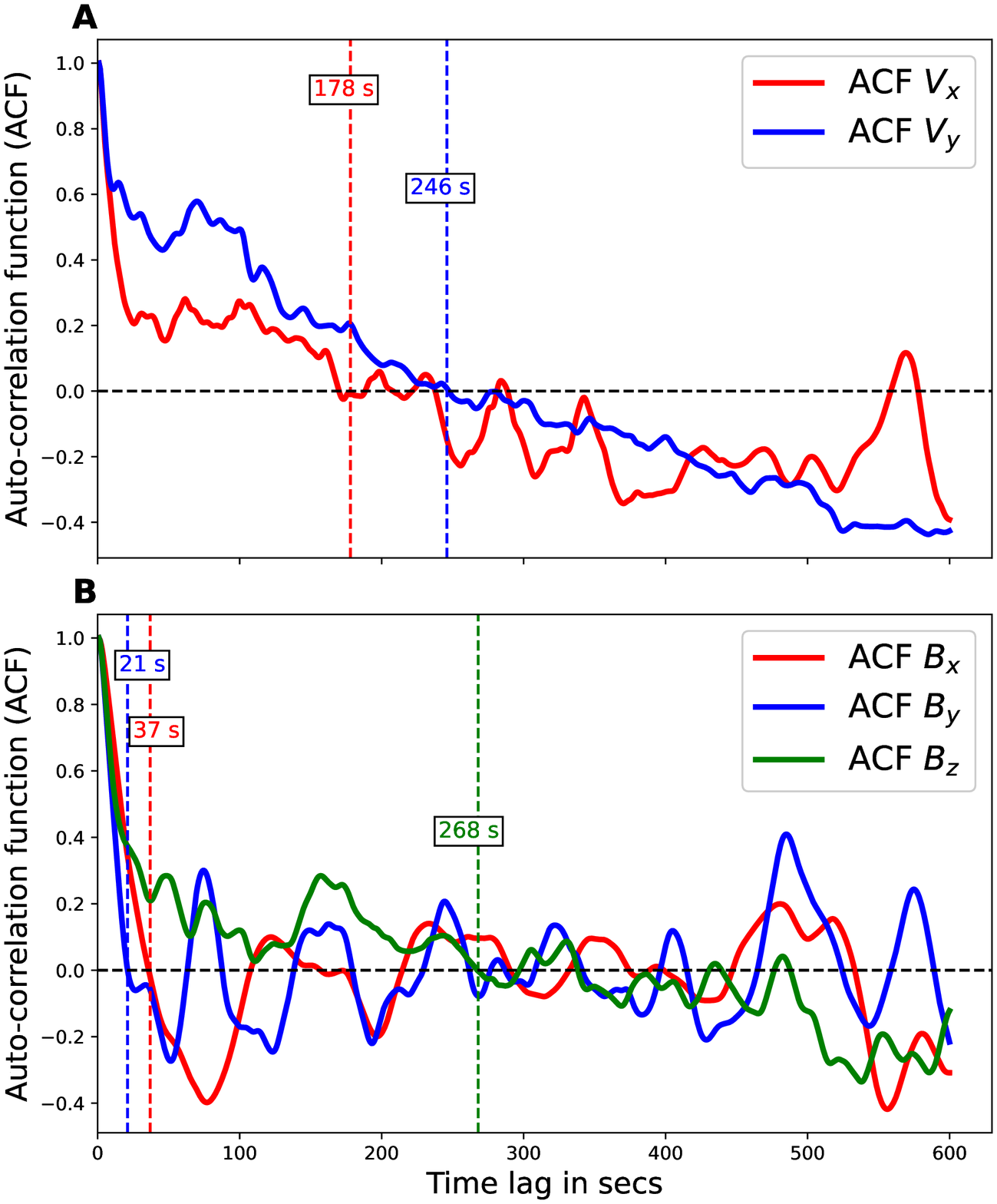}}
	\caption{Panel \textbf{A}: Normalized temporal autocorrelation function for the $x$ and $y$ components of the velocity field at the center of the simulation box (Point A in Figure~\ref{setup}). \textbf{B} : Normalized temporal autocorrelation function for the $x$, $y$, and $z$ components of the magnetic field at the same location. $V_y$ has the highest correlation timescale $ = 246$~s for the velocity components compared to $178$~s for $V_x$. Similarly, $B_z$ has the highest correlation timescale $268$~s, compared to $B_x$ with $37$~s and $B_y$ with $21$~s.}
    \label{acf_velB}
\end{figure}

In contrast to the observations, we can perform autocorrelation analysis for each vector component. All previous analysis of turbulence in prominences \cite[][]{leonardis2012turbulent, freed2016analysis, hillier2017investigating} was carried out on intensity images. In the present study, we make full use of the available field variables (magnetic field magnitude and plasma velocity field magnitude) and quantify their inter-dependency and influence on the turbulent characteristics of our simulation.  
The contribution of each vector component of the velocity and magnetic fields are shown in Figures~\ref{acf_velB} (A and B), respectively. In the case of the velocity field, we see that the $V_y$ has a longer correlation time, 246~s, compared to the $V_x$ component at 178~s. The coherent structures remain for a longer time in the vertical direction and contribute to the increased growth of energy in this direction. This property has been previously observed within prominences (\citet{1976SoPh...49..283E}). The velocity components become increasingly anticorrelated with time. The $B_z$ component of the magnetic field is the largest component of the full magnetic field. Indeed, the correlation time for the $B_z$ component is ten times longer (268~s) than for either the $B_x$ (37~s) or $B_y$ (21~s) components that are of a much smaller magnitude. This confirms how the anisotropic turbulence is influenced by the mean magnetic field orientation (the largest field component is $B_z$ here, perpendicular to the simulation plane). The shorter correlation timescale for poloidal field components, and the longer one for the (mean) field component $B_z$, comparable to the poloidal velocity ones, is in accord with the anisotropy expected.  

\vspace{1cm}
\subsection{Power Spectra}
\begin{figure*}[htbp!]
\centering
   \includegraphics[width=18cm]{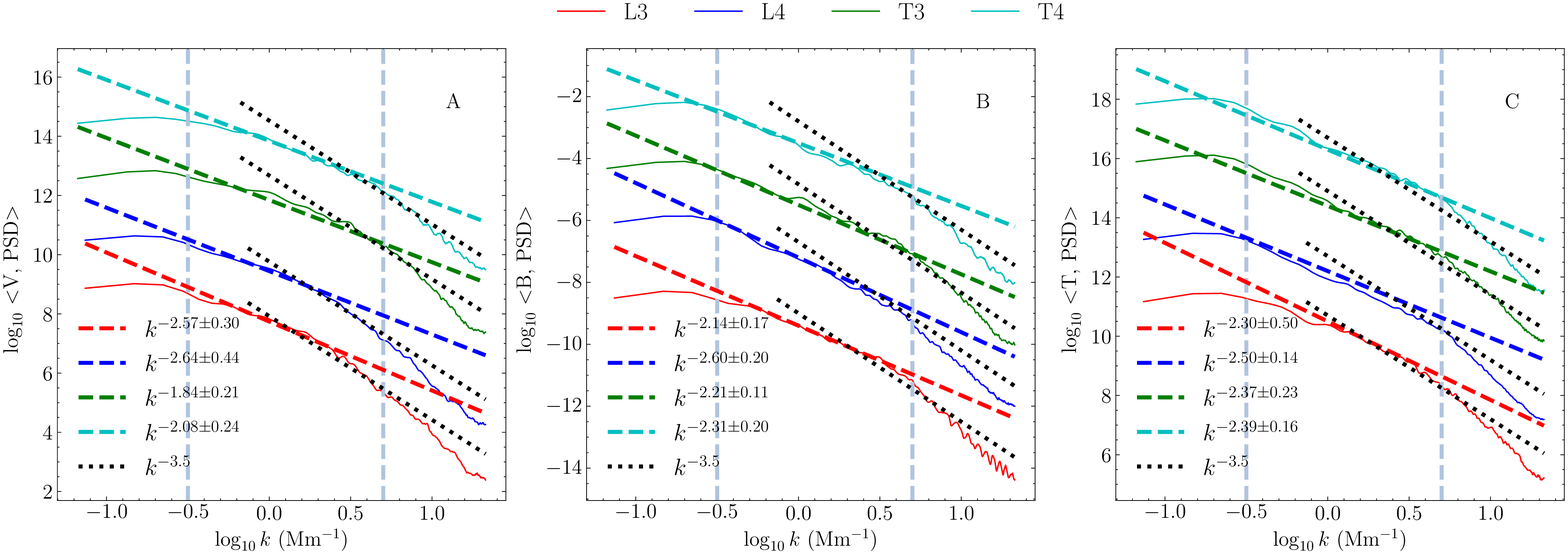}
   \includegraphics[width=18cm]{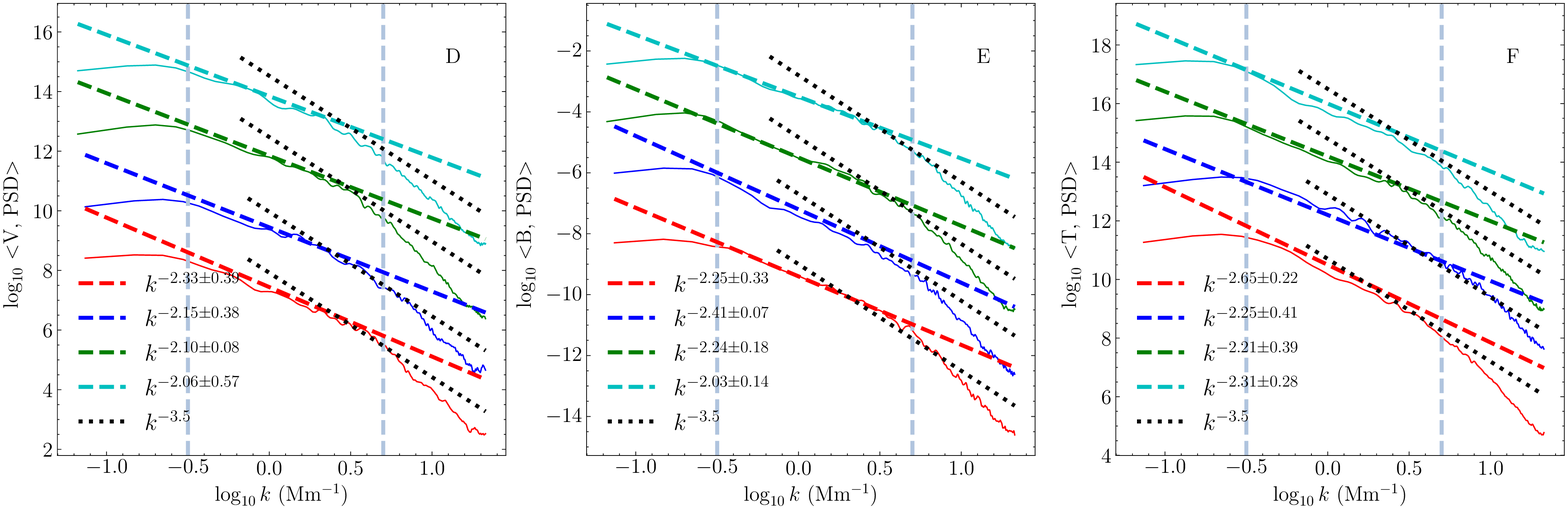}

   \caption{PSD in the wavenumber domain for the longitudinal strips (L3, L4) and transverse strips (T3, T4) are plotted for two ensemble averages of time before the collision of the material with the upper chromosphere boundary representing a more RT instability turbulence (top) and after this collision led to more fully developed turbulence (bottom). All the PSDs are shifted on the $y$-axis for clarity. The dashed colored lines show fits of scalings done for a particular field in one direction between 0.32-3.2~Mm$^{-1}$, as marked by the vertical dashed lines in all panels. There is a secondary steeper scaling -3.5 at larger wavenumbers, as marked by the dotted black lines. }
     \label{PSD v-B}
\end{figure*}

\vspace{0.5cm}
In the present study, the RT instability leads to the development of longitudinally (vertically) dominated structures in the shape of plumes and bubbles, which leads to the development of turbulence by forming initial large-scale disturbances or coherent structures in the solar atmosphere. At a later time $\sim$ 500~s, the cool and dense material begins to collide with the upper chromosphere (transition region) boundary within our domain (the domain extends to lower chromospheric regions), leading to the rise of the structure's transverse (horizontal) motions. This is indicated by the rise in the kinetic energy in the $x$-direction in Figure~\ref{time evol ke by}(A) around that time. An increasing amount of kinetic energy follows after 800~s in both $x$-direction and $y$-direction. Also, visually, this stage resembles a more fully turbulent flow. In a turbulent state, energy flows from the largest scales to the smallest or dissipative ones by a self-similar cascading process. In doing so, turbulence shows a power-law scaling in the power-spectrum, i.e., energy is transported at the same rate throughout the inertial range \citep{kolmogorov1941local}.

We use the Power Spectral Density (PSD) to quantify energy distribution over spatial scales to estimate the power spectrum in a finite data set. If one applies the Fast Fourier Transform (FFT) method to a signal, one may be influenced by noise in the spectra at certain scales. The slope concerning those scales in this method might not have any physical meaning (\citet{bruneau2005comparison}).  To avoid this side-effect, we calculate the PSD of the signal (spatial series) using the Welch periodogram method over the Fast Fourier Transform (FFT) using a Hanning window \citep{welch1967use}. This periodogram helps in smoothing the sample sequence before computing the transform, segmenting the sample sequence, and computing additional estimators that can be averaged using overlapping sample sequences. We divide the transverse signal ($I(r)$) with 1280 points (the number of pixels along each strip) into $G = 1280/M$ segments of $M$ observations each ($M < 1280$), which are allowed to overlap, i.e., $M$ is the size of each segment. If the segments do not overlap, a parameter $S = M$, while for a 50 \% overlap, the parameter $S = M/2$. For each segment, we apply the Hanning window function and compute the windowed periodogram as,
\begin{equation}
    \hat{\phi}_j(r) = \frac{1}{MW_{S}} \Bigg| \sum_{n=1}^M w(n)I_j(n) \exp(-irn) \Bigg|^2 ,
\end{equation}
where $j = 1, \dots, G$, and $W_S$ is the power of the window $w(n)$ given by, $W_S = \frac{1}{M}\sum_{n=1}^M |w(n)|^2$. The average of the windowed periodograms gives us the Welch PSD estimate,
\begin{equation}
    \hat{\phi}_W (r) = \frac{1}{G}\sum_{j=1}^G \hat{\phi}_j (r).
\end{equation}
We follow the same procedure for the longitudinal signal with only 1131 points. Figure~\ref{PSD v-B} shows log-log plots of PSDs for the (L3, L4) strips in the longitudinal and (T3, T4) strips in the transverse direction for the turbulence phase due to RT instability (top) and the fully turbulent phase (bottom). These two phases are defined by taking an ensemble average of snapshots taken before and after falling material collides with the chromosphere.
The ensemble average for the first phase is taken between 558.15-772.83~s and 858.7-1073.38~s for the second phase. The velocity (Panel A and D), magnetic field (Panel B and E), and temperature field (Panel C and F) for the strips are shown in this figure for these two phases. We can see the energy cascades at similar rates for both phases in the field variables. The slope of linear fits to these log-log distributions provide us with the power-law scalings. For both phases, the error fits are calculated for the strips of a field along a direction between a common wavenumber range of $0.32-3.2$~Mm$^{-1}$ in all the panels.

From Figure~\ref{PSD v-B}, we see that the power-law scaling values (also called the spectral index) for all the fields lies in the range between $-1.84$ and $-2.6$, and in Panel A of Figure~\ref{PSD v-B}, the transverse velocity fluctuation is within one $\sigma$ of the famous $-5/3$ spectra for isotropic, homogeneous hydrodynamic turbulence of Kolmogorov. The energy cascading processes on the larger scales in the inertial range is faster compared to such an ideal homogeneous turbulent flow. Of course, the presence of gravity and the perpendicular plane anisotropy likely plays a key role in our inhomogenous and stratified prominence setup, that is, the mixing and the energy dissipation process of the plasma is influenced by the in-plane magnetic field. The other fields also display different and higher scaling than Kolmogorov's $-5/3$. We measure different scaling indices for the larger wavenumber ranges ($>$ 3.2~Mm$^{-1}$) when we do an estimated linear fit, albeit this gives the same average value around $-3.5$ at the end of the inertial range for all the strips (in both directions) for all the fields. The inertial range extends to the wavenumber, where the intermittency effect diminishes, and the unavoidable dissipation (numerical) effects take over in the smallest scales. The larger wavenumbers represent the dissipation scale, with the largest wavenumbers representing the scale of $\sim 23$~km. \vertwo{Comparing the two phases for all the fields, we see shallow spectral indices as we move to more prevalant turbulent behavior throughout the prominence. This indicates that the turbulence level increases once the self-sustaining prominence dissipates as we depart from the RT instability. This behavior can be seen from Figure~\ref{time evol ke by}A, where the rise of the horizontal kinetic energy in comparison to the vertical kinetic energy is observed. The coherent structures in the prominence which is initially dominated in the longitudinal direction are also seen to evolve in the transverse direction after the first phase.}

As far as the observations were concerned, \citet{leonardis2012turbulent} measured the PSDs for the intensity images and yielded slope values of $-2$ for the small wavenumbers ranges between $2.43-4.55$~Mm$^{-1}$ for the longitudinal strips and $2.07-4.61$~Mm$^{-1}$ for the transverse strips. At larger wavenumbers, their spectral index was around $-3$ for transverse strips and $-2.7$ for longitudinal strips. These values for the smaller wavenumbers are similar to that which we find for the velocity, magnetic, and temperature fields. In contrast, for larger wavenumber, our slope values are closer to $-3.5$. \vertwo{Such comparisons are most meaningful for the primitive temperature and velocity fields (although also, but to a lesser degree, for the magnetic field) as their structure and fluctuations influence to first order, albeit in a nontrivial manner, the emergent specific intensity recorded in observations.}

\subsection{Probability Density Functions}

As is the case for many other physical systems, the turbulence within solar prominences is likely chaotic, disorganized, and leading to entirely unpredictable plasma flows. A probabilistic description of turbulence of a field variable $I$ at a position in space is given by its probability density function (PDF), $P(I)$, where $P(I)dI$ is the probability that the variable takes the value between $I$ and $I+dI$ such that it satisfies the normalization condition $\int_{- \infty}^{\infty}P(I)dI = 1$.

For such an analysis, we must create a histogram of the signal, where all the possible signal values are binned into appropriately sized windows. Then the number of counts in each window is divided by the total number of realizations. We obtain the limiting curve of the PDF when the number of realizations increases without bounds as the size of the window goes to zero \citep{george2013lectures}. For this analysis, we concentrate on the turbulence due to the RT instability phase, i.e., at a time of around 558.15~s.
For a stationary Gaussian process, the PDF behaves like a Gaussian, but in the presence of turbulence, the field deviates to a more non-Gaussian process with a heavy-tailed distribution. This is linked to the presence of intermittency or rare events in the signal, i.e., large jumps in values outside of the standard deviation of the mean. This can be seen in the context of hydrodynamic turbulence \citep[][]{anselmet1984high,she1988scale,castaing1990velocity,vincent1991spatial,kailasnath1992probability,mordant2001measurement}, in MHD turbulence \citep[][]{muller2000scaling,biskamp2000scaling,homann2007lagrangian,biskamp2003magnetohydrodynamic}, as well as in solar wind turbulence \citep[][]{burlaga1993intermittent,grauer1994scaling,politano1995model,marsch1997intermittency,	horbury1997structure,sorriso1999intermittency,narita2006wave, koga2007intermittent}.

\begin{figure*}[htbp]
\centering
   \includegraphics[width=12cm]{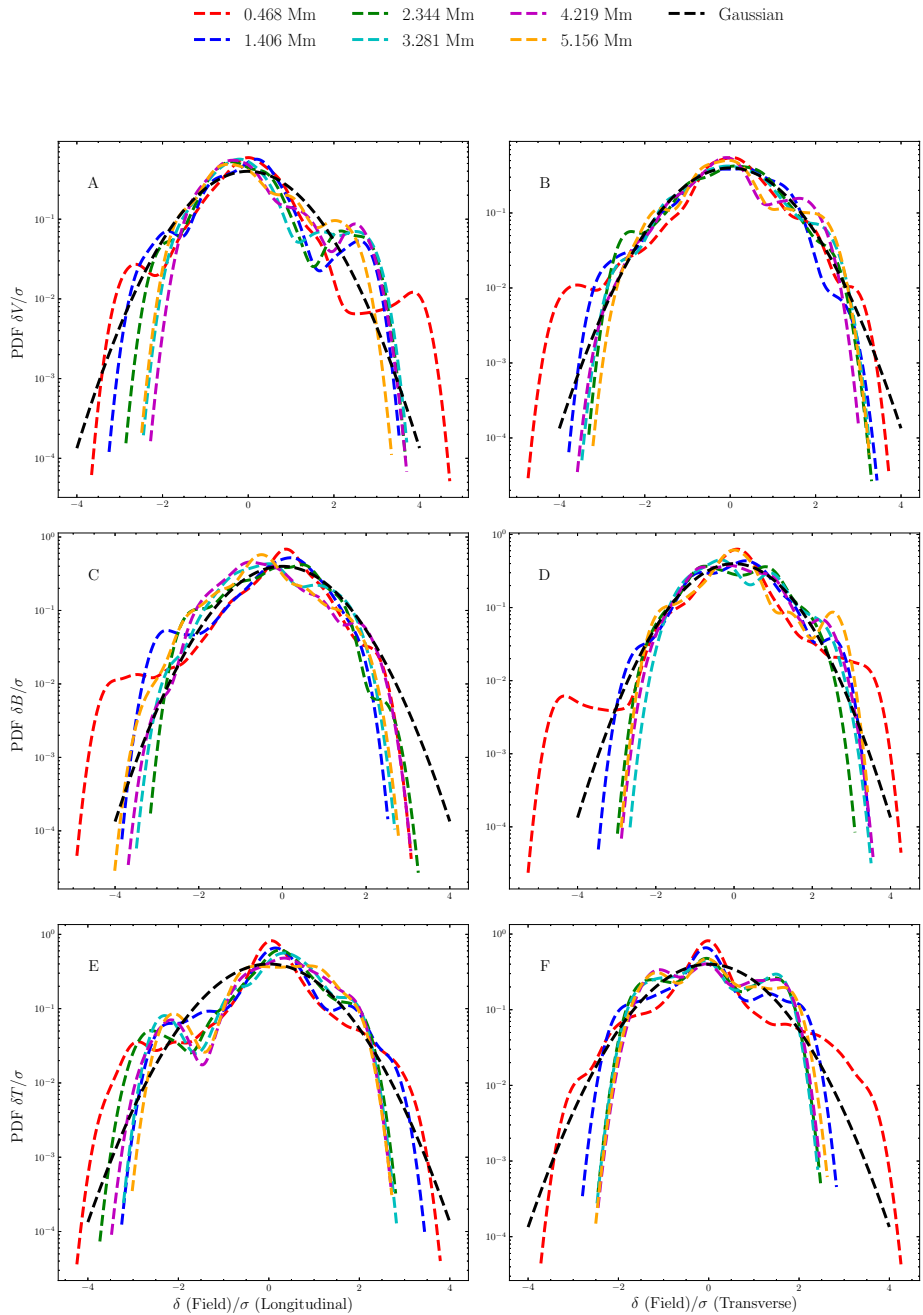}
    \caption{PDF of the velocity (Panel A and B), magnetic (Panel C and D), and temperature field (Panel E and F) increments for L3 in the longitudinal (left) and T3 in the transverse direction (right) for increasing lags (or scales) of L = 0.468~Mm, 1.406~Mm, 2.344~Mm, 3.281~Mm, 4.219~Mm, and 5.156~Mm in the spatial domain plotted against the dashed black line of Gaussian distribution for the reference time 558.15~s.}
    \label{PDF V B}
\end{figure*}

In Figures~\ref{PDF V B} (A and B), the PDF of the velocity field increments $\delta V(r,L)$ is plotted at fixed separations $L$ of $0.468$~Mm, $1.406$~Mm, $2.344$~Mm, $3.281$~Mm, $4.219$~Mm and  $5.156$~Mm in colored lines at time 558.15~s. This is done for the longitudinal L3 strip (A) and in a transverse T3 strip (B), where we normalize to the variance $\sigma$, i.e. $P(\delta V)=\delta V/\sigma$. For reference, the dashed black line is a Gaussian of unit variance plotted to show the degree of intermittency.

In Figures~\ref{PDF V B} (C and D) and (E and F), we plot the same for the magnetic field increments for L3 (C) and T3 strips (D) and temperature field increments for L3 (E) and T3 strips (F) using the same separations $L$, where we again normalize to the variance $\sigma$ of the field. From these Figures~\ref{PDF V B}, we see that the distribution approaches Gaussianity at larger separations $L$. This can also be quantified by the fourth moment of the probability distribution, called Kurtosis, and given by the parameter $K = \langle P(\delta V) \rangle^4 / \sigma^{4}$, which describes the extent of the peak in a distribution. Smaller values (in magnitude) indicate a flatter, more uniform probability distribution \citep{george2013lectures}.
A Gaussian distribution is characterized by a Kurtosis value of 3 and $\overline{K}$ represents the coefficient of excess kurtosis defined as $\overline{K} = K-3$.

From Table~\ref{tab:kurtosis_values}, we see that the excess Kurtosis values for the PDF tend to decrease with increasing lag (separation scale) for the velocity, magnetic, and temperature fields. As such, the PDFs gradually become scale-independent with an increase in scale and eventually approach Gaussianity ($\overline{K}$ becomes close to zero). At such large scales, the fluctuations of the fields are therefore likely to be largely uncorrelated. The smaller scales, on the other hand, show a higher deviation from Gaussianity and indicate a higher rate of intermittent events happening at these scales. From Figure~\ref{PDF V B}, we see the presence of positively skewed values in the longitudinal velocity fields in the small scales describing upward vertical motions, whereas the longitudinal magnetic and temperature fields show more negative values describing downflows in the prominence. \citet{frisch1995turbulence} associates this behavior with intermittency; hence, our results show the existence of intermittency within prominence turbulence.

\begin{table*}[htbp!]
\begin{center}
   \caption{Corresponding to Figure~\ref{PDF V B}, the coefficient of excess Kurtosis values ($\overline{K}$) for the velocity, magnetic, and temperature field increments for L3 and T3 strips at different separation scales $L$ for time 558.15~s.}
   \label{tab:kurtosis_values}
\begin{tabular}{@{}|cccccccl|@{}}
\toprule
\multicolumn{8}{|c|}{\textbf{Coefficient of excess Kurtosis values, ($\overline{K}$)}} \\ \midrule
\multicolumn{2}{|c|}{\textbf{Separations L (in Mm)}} &
  \multicolumn{1}{c|}{0.468} &
  \multicolumn{1}{c|}{1.406} &
  \multicolumn{1}{c|}{2.344} &
  \multicolumn{1}{c|}{3.281} &
  \multicolumn{1}{c|}{4.219} &
  5.156 \\ \midrule
\multicolumn{1}{|c|}{\multirow{2}{*}{V}} &
  \multicolumn{1}{c|}{Longitudinal} &
  \multicolumn{1}{c|}{3.49} &
  \multicolumn{1}{c|}{1.35} &
  \multicolumn{1}{c|}{1.31} &
  \multicolumn{1}{c|}{1.29} &
  \multicolumn{1}{c|}{0.87} &
  0.01 \\ \cmidrule(l){2-8} 
\multicolumn{1}{|c|}{} &
  \multicolumn{1}{c|}{Transverse} &
  \multicolumn{1}{c|}{2.37} &
  \multicolumn{1}{c|}{0.20} &
  \multicolumn{1}{c|}{0.09} &
  \multicolumn{1}{c|}{0.08} &
  \multicolumn{1}{c|}{-0.05} &
  0.2 \\ \midrule
\multicolumn{1}{|c|}{\multirow{2}{*}{B}} &
  \multicolumn{1}{c|}{Longitudinal} &
  \multicolumn{1}{c|}{3.39} &
  \multicolumn{1}{c|}{1.05} &
  \multicolumn{1}{c|}{-0.42} &
  \multicolumn{1}{c|}{-0.53} &
  \multicolumn{1}{c|}{0.08} &
  0.11 \\ \cmidrule(l){2-8} 
\multicolumn{1}{|c|}{} &
  \multicolumn{1}{c|}{Transverse} &
  \multicolumn{1}{c|}{3.57} &
  \multicolumn{1}{c|}{0.18} &
  \multicolumn{1}{c|}{-0.71} &
  \multicolumn{1}{c|}{-0.51} &
  \multicolumn{1}{c|}{-0.21} &
  -0.86 \\ \midrule
\multicolumn{1}{|c|}{\multirow{2}{*}{T}} &
  \multicolumn{1}{c|}{Longitudinal} &
  \multicolumn{1}{c|}{2.96} &
  \multicolumn{1}{c|}{0.88} &
  \multicolumn{1}{c|}{1.52} &
  \multicolumn{1}{c|}{0.94} &
  \multicolumn{1}{c|}{0.81} &
  0.16 \\ \cmidrule(l){2-8} 
\multicolumn{1}{|c|}{} &
  \multicolumn{1}{c|}{Transverse} &
  \multicolumn{1}{c|}{2.00} &
  \multicolumn{1}{c|}{-0.18} &
  \multicolumn{1}{c|}{-0.99} &
  \multicolumn{1}{c|}{-1.06} &
  \multicolumn{1}{c|}{-1.12} &
  -0.9 \\ \bottomrule
\end{tabular}
\end{center}
\end{table*}

\vspace{1cm}
\subsection{Structure functions}
\begin{figure*}[htbp!]
\centering
   \includegraphics[width=15cm]{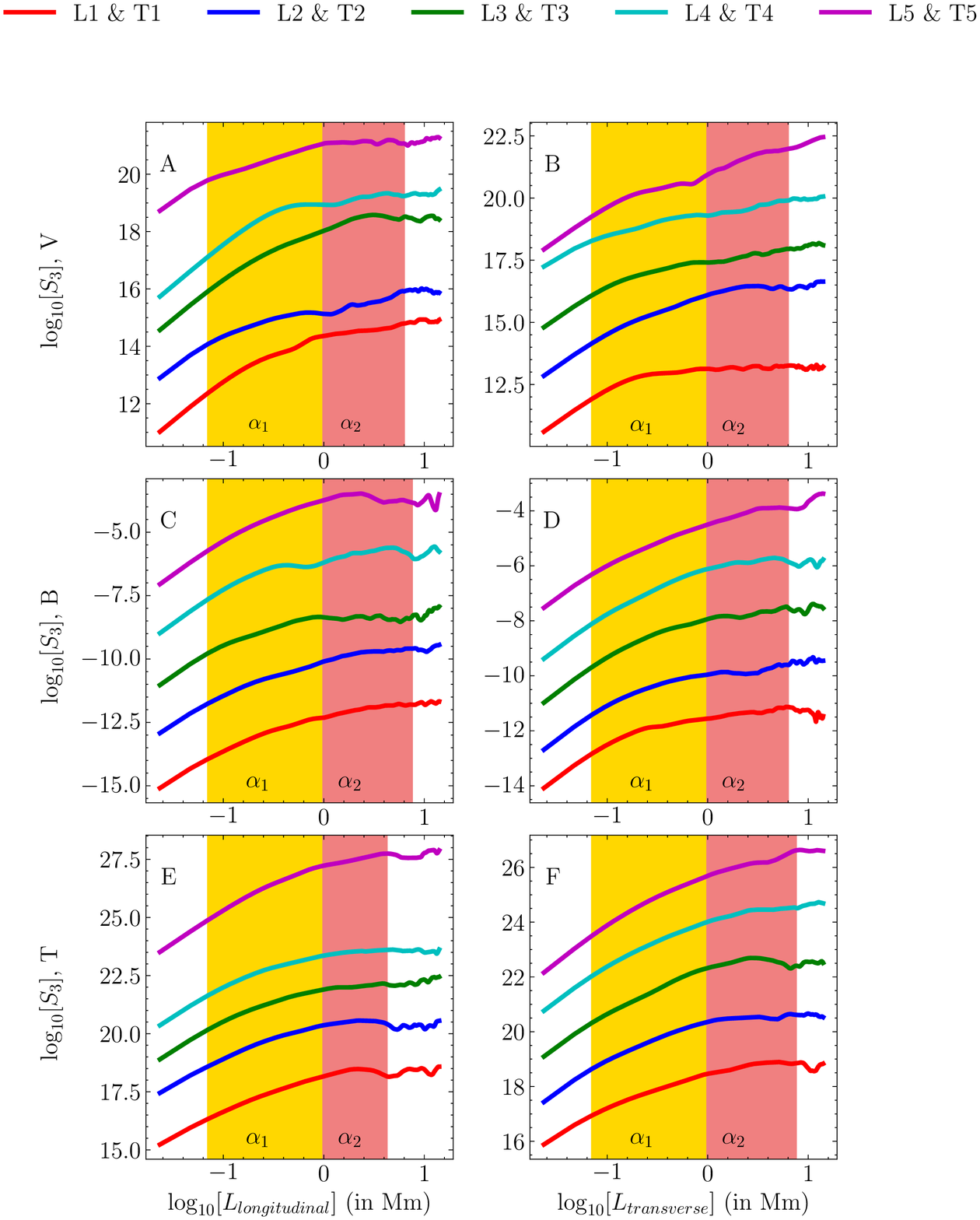}
     \caption{SF (log-log) third order for longitudinal, L1-L5 (left) and transverse strips, T1-T5 (right) of the velocity field (A and B), magnetic field (C and D), and for the temperature field (E and F) at the reference time = 558.15~s. All the SFs are shifted on the $y$-axis for clarity. The rate of intermittency decreases with an increase in scales for A, B, C, D, E, and F. The yellow ($\alpha_1$) and lightcoral ($\alpha_2$) area mark the wavenumber between which the two error fits of power-law exponents are done for the  scales of the individual strips. The values of these fits are given in Table~\ref{tab:alpha values}.}
     \label{SF3}
\end{figure*}

Turbulence within a solar prominence comprises of fluctuations generated by the interaction of coherent structures of all varieties and sizes. Fully developed turbulence shows the property of scale invariance, which is characterized by a sequence of scaling exponents \citep[][]{kolmogorov1941local,frisch1995turbulence}. In the case of scale invariance, it is not possible to identify a predominant scale. Since the power spectrum is not enough for assessing scale invariance in the inertial range, it is necessary to look at higher-order moments. The PDFs qualitatively relate the turbulent dynamics found in prominences, which can be quantitatively realized by calculating the higher moments of these distributions. We use structure functions (SF), which are a measure of the correlation of fluctuations to length scales. An SF of order $p$ of a field $f(r)$ with lag $L$ is defined by,
\begin{equation}
S^p(L) = \langle |(f(r + L)- f(r))|^p\rangle = \int^\infty_{-\infty} | \delta f|^p P(\delta f)d(\delta f) \,,
\end{equation}
where we have taken the absolute value over the distribution and then carried out an ensemble average (we can, in practice, take the mean over the position range in $r$). The SF can describe the characteristics of the turbulent fluctuations in the inertial range, which exhibits a power law function,
\begin{equation}
S^p(L) \sim L^{\zeta(p)},
\label{eq:sfvszeta}
\end{equation}
where $\zeta(p)$ is the scaling exponent. When plotted logarithmically, $\zeta(p)$ gives the slope of the SF. A steeper slope would indicate more intermittency in the distribution. While if the slope changes with the scales, this indicates the change of intermittency as a function of the scale. This behavior is indeed  seen in Figure~\ref{SF3} for the velocity field (in A and B), the magnetic field (in C and D ), and the temperature field (in E and F).

In the case of a self-similar assumption or a fractal (monofractal) process, which is a qualitative measure of having the same behavior of fluctuations at different scales, $\zeta (p)$ scales as a linear function of order p,
\begin{equation}
S^p (L) = C_p L^{\zeta (p)} = C_p L ^{pH} \,,
\end{equation}
where $H$ is the Hurst exponent, ranging between 0 and 1, and $C_p$ is a universal constant. In this case, the relation of the scaling exponent is simply $
\zeta (p) = p H $.
However, fluid turbulence shows a deviation from the linearity of scaling exponents with higher orders of SFs, and this difference with the Kolmogorov self-similar law becomes higher with higher orders of $p$, indicating intermittency. The nonlinearity of the scaling exponent with $p$ thus accounts for multifractal and intermittent behavior.

\begin{table*}[htbp!]
\begin{center}
   \caption{Corresponding to Figure~\ref{SF3}, the two error fits of exponents of power-law are done for the scales between the yellow and lightcoral areas given. The table lists the values for the velocity, magnetic, and temperature field for strips L1-L5 and T1-T5 strips in the longitudinal and transverse direction, respectively, for the time 558.15~s.}
   \label{tab:alpha values}

\begin{tabular}{@{}|cc|c|c|@{}}
\toprule
\multicolumn{2}{|c|}{\textbf{Indices}}                & $\alpha_1$ & $\alpha_2$ \\ \midrule
\multicolumn{1}{|c|}{\multirow{2}{*}{V}} & Longitudinal & $1.35 \pm 0.31$     & $0.46 \pm 0.31$     \\ \cmidrule(l){2-4} 
\multicolumn{1}{|c|}{}                   & Transverse & $0.97 \pm 0.29$    & $0.58 \pm 0.4$     \\ \midrule
\multicolumn{1}{|c|}{\multirow{2}{*}{B}} & Longitudinal & $1.11 \pm 0.31 $    & $0.22 \pm 0.4$    \\ \cmidrule(l){2-4} 
\multicolumn{1}{|c|}{}                   & Transverse & $0.98 \pm 0.28$     & $0.42 \pm 0.33$     \\ \midrule
\multicolumn{1}{|c|}{\multirow{2}{*}{T}} & Longitudinal & $1.39 \pm 0.22$     & $0.33 \pm 0.32$     \\ \cmidrule(l){2-4} 
\multicolumn{1}{|c|}{}                   & Transverse & $1.51 \pm 0.19$     & $0.37 \pm 0.37$     \\ \bottomrule
\end{tabular}%

\end{center}
\end{table*}

\begin{figure*}[htbp]
\centering
   \includegraphics[width=15cm]{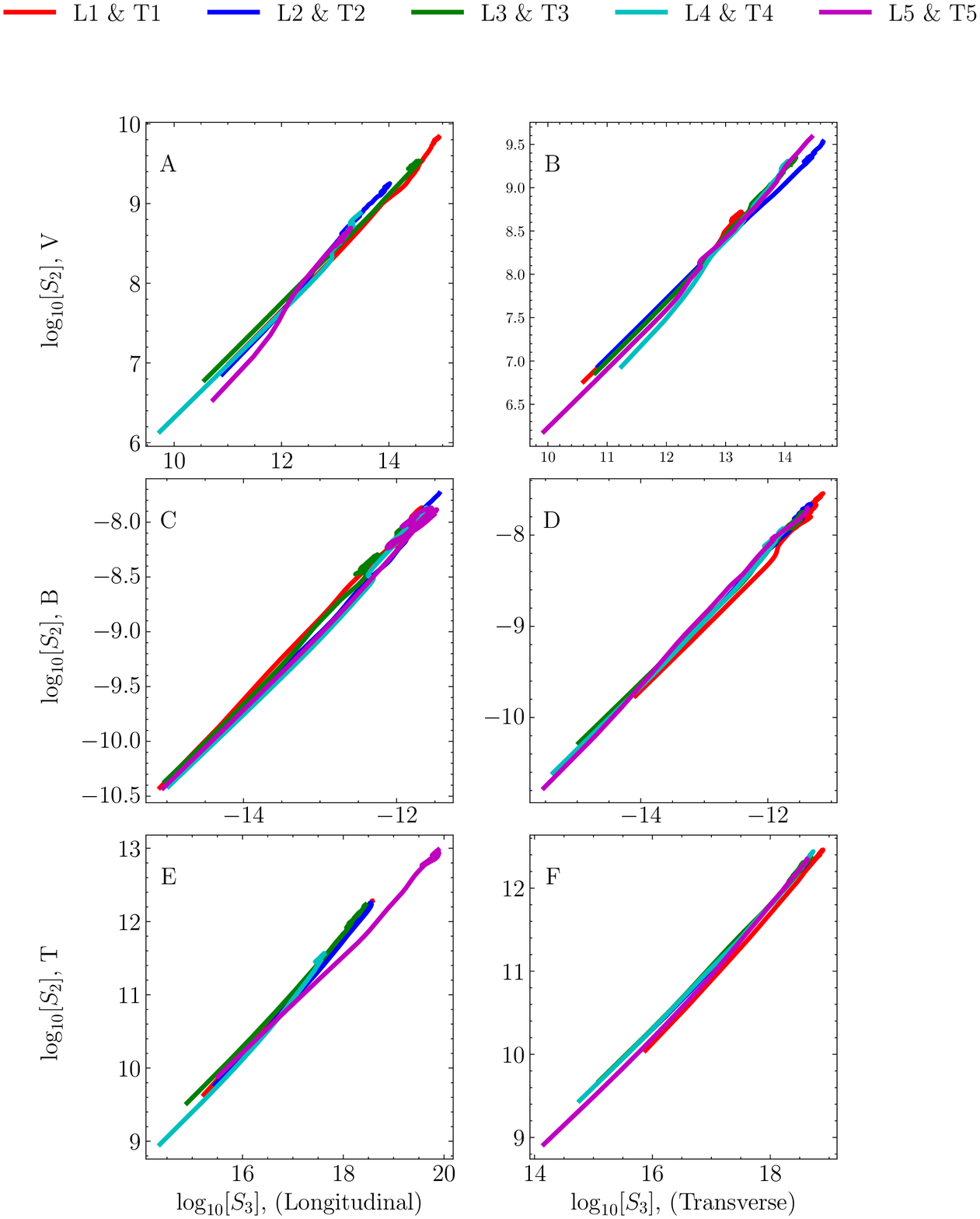}
     \caption{ESS plots (log-log) of SF of order 2 compared to SF of order 3, for all longitudinal strips, L1-L5 (left) and transverse strips, T1-T5 (right) for velocity field (A and B), magnetic field (C and D), and for the temperature field (E and F) at time = 558.15~s. A linear regression line fitting for the entire range of scales in the inertial range gives the slopes, or the scaling exponent $\zeta(2)/\zeta(3)$ is shown in Table~\ref{tab:exponent_values}. $\zeta(2)/\zeta(3)$ values differ from the linear value of 0.66 in each of the cases, implying multifractality.}
     \label{ESS SF2 SF3}
\end{figure*}

Figure~\ref{SF3} shows log-log plots of the third-order SF in the longitudinal direction (left) for L1-L5 strips and in the transverse direction for T1-T5 strips (right) for the velocity (A and B), magnetic (C and D), and temperature field (E and F) at time 558.15~s. The important point to note here is the presence of different slopes in the extended range of scales. We measure the power-law scaling values for the strips and calculate the error fits for the scales in two regions shown in the Figure~\ref{SF3} marked as vertical columns in yellow and lightcoral, respectively. The values of these indices for both strips are given in Table~\ref{tab:alpha values}. In the smallest scales below 0.14~Mm, the scaling is steeper, implying that the intermittency is higher than in the larger scales. For the velocity (longitudinal and transverse) and temperature (longitudinal) fields, the second range of scales spans approximately between 0.14 - 0.96~Mm, as marked in the yellow ($\alpha_{1}$) column, where the slope value is higher in the longitudinal direction compared to the transverse direction. Beyond this range of scales, the third range spans between 0.96 - 6.3~Mm (marked as lightcoral $\alpha_{2}$ column), where the slope values decrease considerably compared to smaller scales for both directions. The third range in the transverse direction of the temperature field spans between 0.96-4.24~Mm. The scales in the range of 6.3~Mm may be the ones driving the upflows and downflows within prominences. The larger scales appear less correlated and may not be able to drive the mean flow arising from the RT instability. The increase of the SF with the scales implies a high correlation of the spatial fluctuations, resulting in the presence of coherent structures in the longitudinal direction. 

In the transverse direction, similar to \citet{leonardis2012turbulent} and \citet{hillier2017investigating}, we find the presence of the knee within the range of 0.2-1.7~Mm for the velocity field and 1.4-5~Mm for temperature field approximately. \citet{leonardis2012turbulent} found this knee range of scales to delimit the crossover between the small-scale turbulence and the large-scale coherent structures. This is also represented in Figure~\ref{PSD v-B}, as the break in the power-law spectrum in the higher wavenumbers of the inertial range where energy is dissipated at a different rate between the small scales and larger scales. This knee could be attributed to an inverse cascade process (typical for true 2D MHD turbulence) which leads to a competition between the energy transfer scales and the intermittent scales responsible for energy dissipation. 

As for the velocity and temperature field, Figure~\ref{SF3} includes the SF of order 3 for the magnetic field in the longitudinal and transverse directions for each of the five strips. We find equivalent features in this case; however, the SFs in the longitudinal direction of the magnetic field are steeper and highly intermittent in the second range of scales, i.e., 0.14 - 0.96~Mm. As we move to higher scales in the magnetic field, the third range spans between 0.96-7.57~Mm in the longitudinal direction and 0.96-6.3~Mm in the transverse direction. The slope values for the third range of scales are lower than that of smaller scales for the magnetic fields, which we have already noted for the velocity and temperature fields. We also see the presence of the knee in the transverse direction of the magnetic field, lying approximately between 0.64 - 3.8~Mm.

\begin{table*}[htbp!]
\begin{center}
   \caption{Corresponding to Figure~\ref{ESS SF2 SF3}, the scaling exponent values $\zeta(2)/\zeta(3)$ for the velocity, magnetic, and temperature field for strips L1-L5 and T1-T5 strips in the longitudinal and transverse directions respectively, for time 558.15~s}
   \label{tab:exponent_values}

\begin{tabular}{@{}|ccccccc|@{}}
\toprule
\multicolumn{7}{|c|}{\textbf{Scaling exponent}, $\zeta(2)/\zeta(3)$} \\ \midrule
\multicolumn{2}{|c|}{\textbf{Strip Number}} &
  \multicolumn{1}{c|}{1} &
  \multicolumn{1}{c|}{2} &
  \multicolumn{1}{c|}{3} &
  \multicolumn{1}{c|}{4} &
  5 \\ \midrule
\multicolumn{1}{|c|}{\multirow{2}{*}{V}} &
  \multicolumn{1}{c|}{Longitudinal} &
  \multicolumn{1}{c|}{0.83} &
  \multicolumn{1}{c|}{0.74} &
  \multicolumn{1}{c|}{0.72} &
  \multicolumn{1}{c|}{0.84} &
  0.81 \\ \cmidrule(l){2-7} 
\multicolumn{1}{|c|}{} &
  \multicolumn{1}{c|}{Transverse} &
  \multicolumn{1}{c|}{0.74} &
  \multicolumn{1}{c|}{0.58} &
  \multicolumn{1}{c|}{0.72} &
  \multicolumn{1}{c|}{0.89} &
  0.80 \\ \midrule
\multicolumn{1}{|c|}{\multirow{2}{*}{B}} &
  \multicolumn{1}{c|}{Longitudinal} &
  \multicolumn{1}{c|}{0.76} &
  \multicolumn{1}{c|}{0.81} &
  \multicolumn{1}{c|}{0.74} &
  \multicolumn{1}{c|}{0.80} &
  0.65 \\ \cmidrule(l){2-7} 
\multicolumn{1}{|c|}{} &
  \multicolumn{1}{c|}{Transverse} &
  \multicolumn{1}{c|}{0.76} &
  \multicolumn{1}{c|}{0.73} &
  \multicolumn{1}{c|}{0.74} &
  \multicolumn{1}{c|}{0.76} &
  0.73 \\ \midrule
\multicolumn{1}{|c|}{\multirow{2}{*}{T}} &
  \multicolumn{1}{c|}{Longitudinal} &
  \multicolumn{1}{c|}{0.80} &
  \multicolumn{1}{c|}{0.82} &
  \multicolumn{1}{c|}{0.81} &
  \multicolumn{1}{c|}{0.94} &
  0.74 \\ \cmidrule(l){2-7} 
\multicolumn{1}{|c|}{} &
  \multicolumn{1}{c|}{Transverse} &
  \multicolumn{1}{c|}{0.84} &
  \multicolumn{1}{c|}{0.80} &
  \multicolumn{1}{c|}{0.80} &
  \multicolumn{1}{c|}{0.81} &
  0.83 \\ \bottomrule
\end{tabular}
\end{center}
\end{table*}

Similar to \citet{leonardis2012turbulent}, we found that the SFs do not follow the power law scaling given by Eq.~(\ref{eq:sfvszeta}) which describes a linear relationship of the scaling exponent $\zeta (p)$ with $p$. This may be attributed to systems limited by finite-range turbulence effects \citep[][]{dubrulle2000finite,bershadskii2007beyond} where turbulence is not fully developed. As given by the theory of Extended Self-Similarity (ESS) \citep[][]{benzi1993extended,carbone1996solar}, we can characterize the turbulence in the prominence using a set of scaling exponents $\zeta(p)$, by comparing SFs of a different order, $p$ against $q$ which is given by,
\begin{equation}
S^p(L) = (S^q(L) )^{\zeta (p)/ \zeta (q)} \,,
\end{equation}
We derive the ratio of the scaling exponents of different orders by calculating the slope of the line from the log-log plot of $S^p$ versus $S^q$. We plot the same for $S^2$ versus $S^3$ in the log-log scale to obtain the ratio of $\zeta (2)/ \zeta (3)$ from Figure~\ref{ESS SF2 SF3} for the velocity field (in A and B), magnetic field (in C and D), and temperature field (in E and F) respectively at time 558.15~s, panels on the left (right) present the results for the five longitudinal (transverse) strips. Using linear regression fits for the curves over the whole range of scales in the inertial range, we obtain the values for the corresponding ratio of scaling exponents  $\zeta (2)/ \zeta (3)$ for the longitudinal and transverse strips. The ratio $\zeta (2)/ \zeta (3)$ is shown for the corresponding direction and field in Table~\ref{tab:exponent_values}. They are found to be different from the linear value of $2/3 = 0.66$  for each of the strips and imply the non-linear form of the scaling exponents $\zeta (p)$.  We thus find that the characteristics of turbulence in our simulated prominences due to the RT instability indicate multifractal behavior and show a high degree of intermittency in the smaller scales. 

\begin{figure*}[htbp!]
\centering
   \includegraphics[width=15cm]{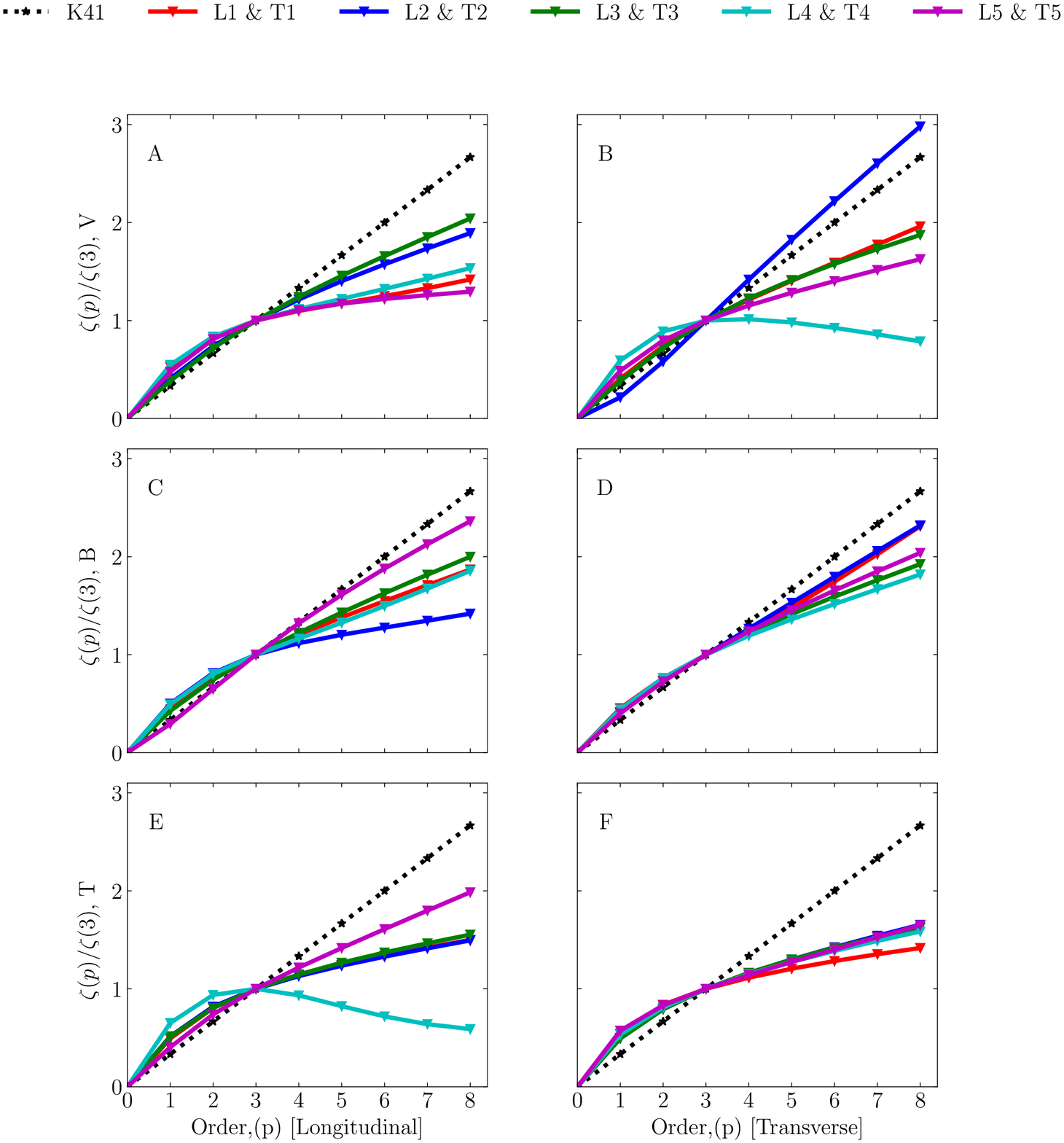}
     \caption{The scaling exponent $\zeta(p)/\zeta(3)$, inferred from the
different slopes from Figure~\ref{ESS SF2 SF3} for the longitudinal (left) and transverse strips (right) for the velocity (A and B), magnetic (C and D), and temperature field (E and F) are plotted versus order $p$, which goes from one to eight at time 558.15~s. The black dotted line here represents the K41 self-similarity line or linear line ($p/3$).}
     \label{SF scaling velB}
\end{figure*}

We see from Figure~\ref{SF3} that the small scales of both the velocity, magnetic, and temperature fields are highly intermittent compared to the larger scales, and as such, with higher orders of SF, i.e., increasing $p$, would give more weight to extreme and rare events in a turbulent flow. Table~\ref{tab:exponent_values} clearly gives the value of multifractality from the ESS theory for order two compared to order three. In comparison, Figures~\ref{SF scaling velB} shows the scaling exponents $\zeta(p)/\zeta(3)$ for all the strips of the velocity (A and B), magnetic (C and D), and temperature (E and F) fields respectively, against all orders from one to eight for longitudinal (transverse) in the left (right) panels for the two reference time. The black line shows the K41 \citep{kolmogorov1941local} self-similarity law which behaves linearly with respect to the orders of SFs. For all the strips in the longitudinal and transverse direction of the velocity and the magnetic field, the scaling exponents deviate from the \vertwo{linear self-similarity line}, and as such, the turbulence in prominence clearly shows intermittency behavior with a higher degree of non-linearity as we focus on the smaller scales and increase the order of SFs. The exponents from higher-order structure functions are prone to errors as with the increase of the order $p$, it increasingly affects the $S^p(L)$ in the presence of spurious large fluctuations. \citet{1997PhRvE..55.2789L} and \citet{2003ApJ...583..308P} used a procedure to determine the highest significant order by calculating the peak of the histogram of $| \delta f|^p$ which
occurs for a value of $\delta f$ that must be represented by a significant
number of points in the PDF of $\delta f$. \citet{hillier2017investigating} used an empirical way to calculate the highest order applicable with a reasonable noise for the study of higher-order structure functions. So, we adopt caution on the exponent values for orders more than 6 in our case, as they might be susceptible to errors.

\section{Conclusion}\label{sec-conc}
In this paper, we use ideal MHD simulations in quiescent prominence conditions to analyze the turbulent characteristics obtained from the simulation data, which transitioned starting from a phase of the linear RT stage to a non-linear, and, finally a turbulent stage at a much later time. The formation of coherent structures from falling bubbles and rising pillars into more organized structures predominantly in the vertical direction reveals the role of turbulence in mixing the stratified prominence flow. Earlier observations from Hinode SOT gave a first indication of the presence of turbulence within the prominences and the influence of characteristic scales on driving the turbulence. The analysis from observed intensity images was made under certain assumptions, such as assuming that moving structures in the images follow the flow, acting as markers or passive scalars for the plasma dynamics. They were restricted to line-of-sight measurements. In our study, we have presented the results of an ideal MHD simulation representing a solar prominence within a $30 \times 30$~Mm domain. Once the dynamics within the simulation entered a turbulent state, analysis of these snapshots revealed and quantified the turbulent behavior of the magnetic and velocity field fluctuations. 

\vertwo{We find a longer correlation time for the $V_y$ component compared to the $V_x$ component indicating the vertical flows are more sustained, as remarked upon within observations.} We investigated the power-law scaling that was evident from the PSDs of the velocity and the magnetic fields so as to ascertain the scaling at which energy was cascading from larger to smaller scales. We found these power-law scaling values to be steeper than the Kolmogorov's $k^{-5/3}$ scaling, in the inertial range for the velocity, temperature, and magnetic fields. Additionally, there is a secondary scaling of $k^{-3.5}$, identifiable for all the fields at the higher wavenumbers of the inertial range. This is suggestive of the turbulence within the simulation experiencing numerical dissipation in those wavenumbers. \vertwo{We observe shallow spectral indices for the fields as we move from the RT instability-generated prominence which dissipates at a later time to a higher turbulent regime representing higher turbulence behavior.} As the fitting used to determine the spectral index is only performed over one decade, it can be very hard to relate the turbulent characteristics to the theoretical work of Kolmogorov or Iroshnikov-Kraichnan, so caution should be exercised. When we check for intermittency, the PDFs of the velocity and magnetic field increments display a constant increase in spatial separations, indicating a departure from Gaussianity. The Kurtosis of the distributions further indicates the magnitude of the non-Gaussianity and intermittency. As we increase the separations, the distributions return to an approximate Gaussian, and the values of excess Kurtosis approach zero. At the larger spatial scales, the distribution loses its coherency and becomes more Gaussian, i.e., they are less correlated. We note a more non-Gaussian distribution with higher intermittency for the smaller scales. We explore this further with the use of SFs to analyze their behavior within the inertial range and their role in developing turbulence across scales. The third-order SF illustrates different slope values for the scales in the inertial range, which could be classified into three ranges. The steeper the slope, the higher the degree of intermittency present. The first range of smallest scales till 0.14~Mm for all the fields are steeper in the longitudinal direction compared to the transverse direction, which is the same for the second range of scales between 0.14 - 0.96~Mm for the velocity, temperature, and magnetic fields in both directions. Similar to observations \citep{leonardis2012turbulent,freed2016analysis,hillier2017investigating}, we note a break in the SFs across the scales of around 2 - 5~Mm for the velocity, magnetic, and temperature fields in the transverse direction, which was previously interpreted as a physical limit in the ability of small scale turbulence to cross over into the larger scales. The longitudinal velocity increments also show a higher degree of correlation, constituting the domination of flow in the vertical direction, which is also seen from the observations. Finally, the degree of multifractal turbulence is shown with the ratio of scaling exponents $\zeta(2)/\zeta(3)$ whose values differ from the self-similar assumption of linearity or fractal turbulence, i.e., 0.66.

In the present numerical study of prominence formation due to RT instability, we studied the development of prominence features within a 2.5D setup with the mean magnetic field directed into the plane. We use an ideal MHD prescription, and hence the scales where numerical dissipation matters will be strongly sensitive to the chosen numerical resolution. We plan to address true dissipation effects in follow-up work, where we will use fully converged, resistive MHD studies at numerically affordable (i.e. below actual) magnetic Reynolds numbers. There are limitations with the current choice of 2.5D setup, and this might not be an accurate representation of the 3D turbulence that can develop in the solar prominence, but our numerical simulation presents a strong case that RT instability is one of the main contributors to turbulence in prominences. In our 2.5D setup, plasma flow is heavily influenced by the out-of-plane mean-magnetic field, so we urgently need a fully 3D counterpart for this study. Note that the chosen initial angle of the magnetic field to the simulated poloidal plane influences the linear instability properties and the ensuing turbulence. In our setup, magnetic tension is initially zero, since the 2.5D setup has only magnetic pressure support for the prominence. Magnetic tension in the simulated plane does play a role within the further evolutions, as shown in Figure~\ref{by_evolution}. But, of course, future 3D work can start from more realistic, magnetic-tension-supported prominence models where the dense matter initially rests in upwardly concave dips.

Furthermore, in an ideal MHD setup, we do not consider the dissipative scales, which have an effect both on the linear and non-linear phase of the ensuing numerical magnetic reconnection within the simulation. Magnetic reconnection is responsible for the intermittent events of heating and energy dissipation. \citet{hillier2021jets}, using observations from the Interface Region Imaging Spectrograph (IRIS), investigated the occurrence of magnetic reconnection events through the ejection of bi-directional blobs along the current sheet developed within an observed prominence. We will look at such reconnection events in a numerical resistive model follow-up study under prominence conditions.
The magnetic Prandtl number is given by $Pm = Rm/Re = \nu / \eta$, where $Rm$ is the magnetic Reynolds number, $Re$ is the Reynolds number, $\nu$ is the kinematic viscosity, and $\eta$ is the magnetic diffusivity. This parameter determines which scale is larger: the viscous dissipation scale or the resistive dissipation scale. In the solar atmosphere, the value is very low. Thus, in order to conclude more about the effects of these scales, we need to include the resistive MHD model within our simulation setup. Moreover, our model lacks additional physics to the system, such as the Hall effect, partial-ionization (ambipolar diffusion), different ion-electron temperatures, radiative losses, and thermal conduction. The inclusion of any combination of these effects would extend our understanding of turbulence within solar prominences. This is the focus of future endeavors.

\begin{acknowledgements}
This work was supported by the ERC Advanced Grant PROMINENT, and an FWO grant G0B4521N. This
project has received funding from the European Research Council (ERC) under
the European Union’s Horizon 2020 research and innovation programme (grant
agreement No. 833251 PROMINENT ERC-ADG 2018). This research is further supported by Internal funds KU Leuven, through the project C14/19/089 TRACESpace.
\end{acknowledgements}


\bibliographystyle{aa} % style aa.bst
\bibliography{main_aa} % your references Yourfile.bib

\end{document}